\documentclass[12pt]{iopart}

\usepackage{graphicx}
\usepackage{color}
\usepackage{amsmath}
\usepackage[dvipsnames]{xcolor}

\begin{document}

\title {Emerging new phases in correlated Mott insulator Ca$_2$RuO$_4$
}

\author{Giuseppe Cuono$^1$, Filomena Forte$^2$, 
Alfonso Romano$^{2,\, 3}$ and Canio Noce$^{2,\,3}$\footnote{Corresponding author}}

\address{$^1$ CNR-SPIN,
c/o Università degli Studi "G. D'Annunzio", 66100 Chieti, Italy}
\address{$^2$ {CNR-SPIN, c/o Università degli Studi di Salerno - Via Giovanni Paolo II, 132 - 84084 - Fisciano (SA), Italy}}
\address{$^3$ {Dipartimento di Fisica "E R Caianiello", Università degli Studi di Salerno - Via Giovanni Paolo II, 132 - 84084 - Fisciano (SA), Italy}}
\ead{giuseppe.cuono@spin.cnr.it}
\ead{filomena.forte@spin.cnr.it}
\ead{alromano@unisa.it}
\ead{cnoce@unisa.it}

\vspace{10pt}
\begin{indented}
\item[]May 2024
\end{indented}

\begin{abstract}
The Mott insulator Ca$_2$RuO$_4$ is a paradigmatic example among transition metal oxides, where the interplay of charge, spin, orbital, and lattice degrees of freedom leads to competing quantum phases.
In this paper, we focus on and review some key aspects, from the underlying physical framework and its basic properties, to recent theoretical efforts that aim  to trigger unconventional quantum ground states, using several external parameters and stimuli.
Using first-principle calculations, we demonstrate that Ca$_2$RuO$_4$ shows a spin splitting in the reciprocal space, and identify it as an altermagnetic candidate material. The non relativistic spin-splitting has an orbital selective nature, dictated by the local crystallographic symmetry.
Next, we consider two routes that may trigger exotic quantum states. The first one corresponds to transition metal substitution of the 4$d^4$ Ru with isovalent 3$d^3$ ions. 
This substitutional doping may alter the spin-orbital correlations favoring the emergence of negative thermal expansion. The second route explores fledgling states arising in a non-equilibrium steady state under the influence of an applied electric field. We show that the electric field can directly affect the orbital density, eventually leading to strong orbital fluctuations and the suppression of orbital imbalance, which may, in turn, reduce antiferromagnetism. These aspects suggest possible practical applications, as its unique properties may open up possibilities for augmenting existing technologies, surpassing the limitations of conventional materials.

\end{abstract}

\section{Introduction}

The interplay of spin, orbital, and lattice degrees of freedom in the $t_{2g}$ manifold, together with the role played by intermediate/strong electronic correlations, has emerged as a crucial aspect in understanding and predicting the intricate physical properties of numerous transition metal oxides. Indeed, by intertwining these fundamental properties, a plethora of intriguing phenomena can emerge, from exotic magnetic behaviors to novel electronic phases.

A widely studied class of systems where these effects are particularly evident is that of the ruthenate compounds belonging to the Ruddlesden-Popper series having the general formula A$_{n+1}$Ru$_n$O$_{3n+1}$ with A = Ca or Sr. This can largely be ascribed to the presence in these systems of ruthenium, which is a transition metal with a partially filled 4$d$ shell. Since in a solid $d$ orbitals remain fairly localized around the ion to which they belong, they tend to give rise near the Fermi level to narrow bands where the electronic correlations are generally nonnegligible. However, {in materials containing atoms with partially filled 4d shells,} the larger extension of the 4$d$-orbitals compared to the 3$d$ ones makes the value of the intra-orbital Coulomb repulsion not very high and usually comparable {with the width of the bands developing around the Fermi level.} This is accompanied by a stronger hybridization between $d$ and $p$ orbitals, so that these systems tend to be on the border between metallic and insulating behavior and/or on the verge of long-range magnetic order. 
In addition, spin-orbit coupling also plays a relevant role, since it acts on an energy scale comparable to the other energy scales of the system. The observed physical properties are thus the result of a complex interplay of Coulomb interactions, hybridization, spin-orbit splitting, and crystal field effects, so that small perturbations such as slight pressure-induced lattice distortions, application of a magnetic field, small variations in the doping concentration, etc., can easily induce drastic changes in the ground state.

In this context, much attention in the last years has in particular been devoted to Ca$_2$RuO$_4$.
Synthesis and characterization of this system began in the late 1990s~\cite{Nakatsuji97a,Nakatsuji97b}, when the extensive studies previously conducted on Sr$_2$RuO$_4$ motivated researchers to extend their investigations to other related $4d$ transition metal oxides. Despite the similarity of undoped Ca$_2$RuO$_4$ and Sr$_2$RuO$_4$, many differences in their electronic properties were soon identified.
While Ca$_2$RuO$_4$ in the ground state is a Mott insulator exhibiting antiferromagnetic order, Sr$_2$RuO$_4$ is a good Fermi liquid approximately below 25 K, becoming an unconventional superconductor below $T_c=1.5\,$K. The evolution of the ground-state properties in going from one system to another has been investigated in Ref.~\cite{Nakatsuji00PRL}, where the authors were able to control carefully the relative concentrations of Sr and Cr in Ca$_{2-x}$Sr$_x$RuO$_4$.

Research activity on Ca$_2$RuO$_4$ to date is still very intense and has recently led to new results of particular interest. Some of them are presented and discussed here. In particular, after a short summary of the structural and electronic properties, we discuss the emergence in Ca$_2$RuO$_4$ of a new form of magnetism, the so-called altermagnetism, such that the magnetization in the real space is zero due to antiparallel orientation of spins, like in antiferromagnets, but the band structure is characterized by non-relativistic spin-splitting, like in ferromagnets~\cite{Smejkal22,Smejkal22b}. 

Then we present another fascinating phenomenon that characterizes the behavior of Ca$_2$RuO$_4$, that is, the negative thermal expansion (NTE). It is a counter-intuitive phenomenon observed in certain materials, where the volume contracts rather than expands upon heating. This unique property, observed in various compounds and alloys, challenges our conventional understanding of thermal behavior because, traditionally, materials tend to expand when heated due to increased thermal vibrations. Nonetheless, there exist many diverse examples of systems where this phenomenon can be ascribed to electronic mechanisms rather than to the canonical structural one. It was recently shown~\cite{Brzezicki23} that in Ca$_2$RuO$_4$ electron correlations can lead to NTE, this effect being triggered by inverse crystal-field potential acting on spin-orbital degrees of freedom.

Finally, we present discuss another attempt to control the insulated spin-orbital ordered state of Ca$_2$RuO$_4$, via the use of  out-of-equilibrium driving forces. In particular, the application of electric fields can govern the electronic states directly, for example it can affect an existing directional order of the orbitals, and may in turn alter the magnetic and conducting state. In the case of Ca$_2$RuO$_4$, recent experiments showed that in response to a static electric current, the insulating state exhibits non-equilibrium steady states with anomalous electrical behavior~\cite{Nakamura2013,Okazaki2013} and stripe patterns~\cite{Zhang2019,Gauquelin23}, indicating that such low-conducting phase can be quite different from the thermal-driven metallic state. In this work, we concentrate on how an external electric field may affect the orbital order in Ca$_2$RuO$_4$ and we theoretically simulate the charge and orbital reconstruction induced by the applied electric field.

The paper is organized as follows. In the next section, we review the structural and electronic properties of Ca$_2$RuO$_4$, while the following three sections are devoted to specific new phases emerging in this system. Specifically, in Section 3, we explain why this compound can support an altermagnetic state. Section 4 provides a theoretical description of the NTE effect exhibited by Ca$_2$RuO$_4$. In Section 5, we demonstrate how the electric field can affect the orbital properties of this compound. Finally, in the last section, we present the conclusions and discuss some specific issues that may be addressed in the near future.

\section{Structural and electronic properties}

\begin{figure}[!bth]
\begin{center}
\includegraphics[width=0.4\textwidth]{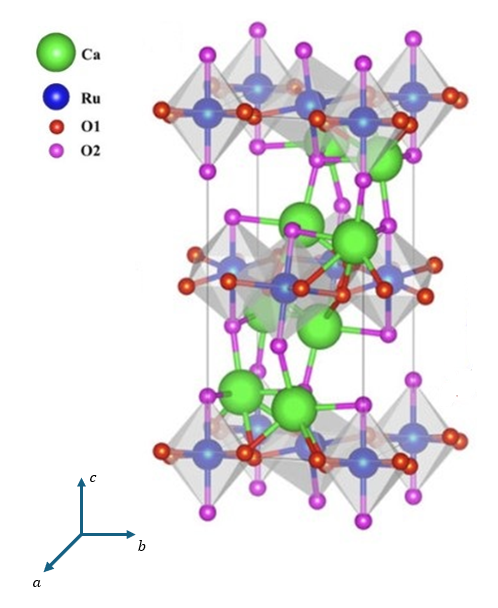}
\end{center}
\caption{Crystal structure of Ca$_2$RuO$_4$.}
\label{crystal}
\end{figure}

\begin{figure}[!bth]
\begin{center}
\includegraphics[width=0.6\textwidth,height=0.7\textheight]{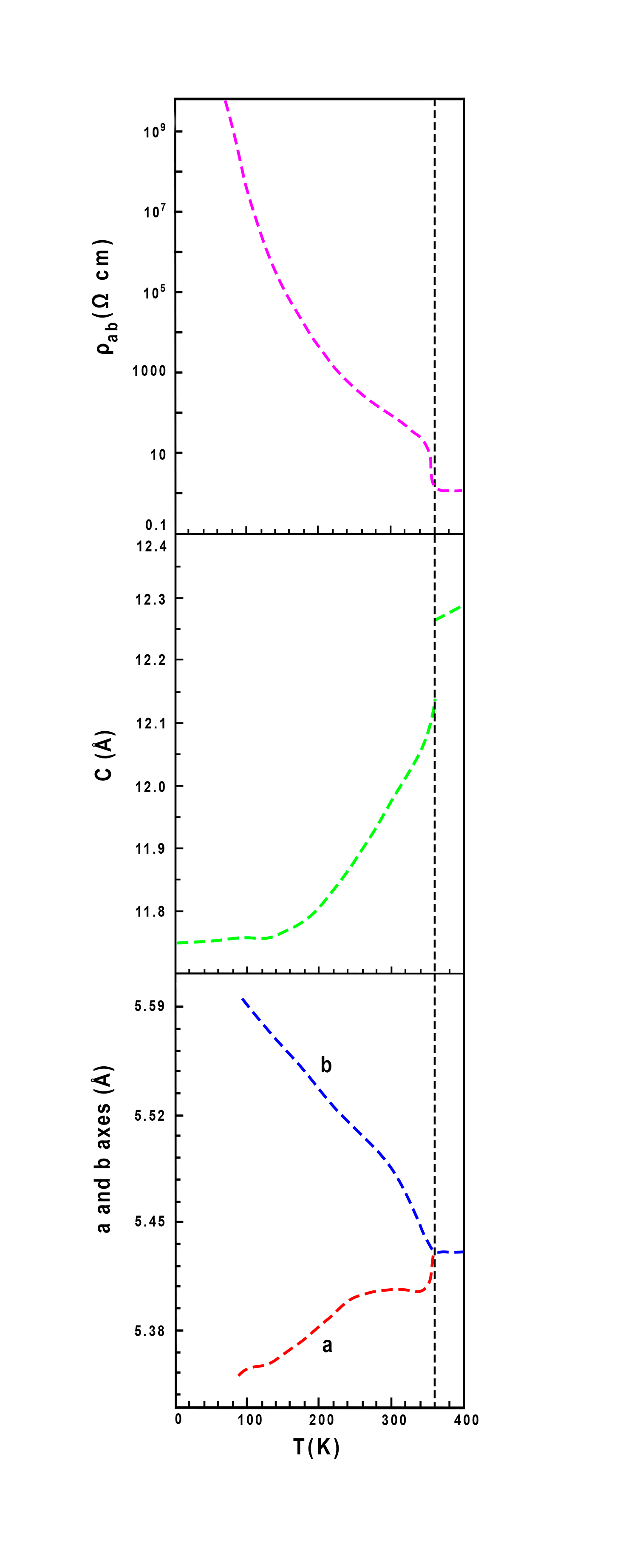}
\end{center}
\caption{Upper panel: temperature dependence of the $ab$-plane resistivity. Middle panel: temperature dependence of the $c$-axis lattice constant. Lower panel: temperature dependence of the $a$- and $b$-axis lattice constants. The vertical dotted line corresponds to  metal-to-insulator transition temperature $T_{\rm MIT}=357\,$K (adapted with permission from Ref.~\cite{Alexander99}, copyrighted by the American Physical Society).}
\label{rho-c_axisT}
\end{figure}

Diffraction methods using X-rays, suited for superficial analysis or thin samples, or neutrons, having high penetration depth and thus more suited for bulk analysis, have shown that the crystal structures of Ca$_2$RuO$_4$ and Sr$_2$RuO$_4$ exhibit important differences, despite having both as fundamental structural units RuO$_6$ octahedra arranged in corner-shared planes alternated by layers containing the cations Ca or Sr. While Sr$_2$RuO$_4$ has a I$4/mmm$ tetragonal symmetry with a unit cell containing two Ru atoms, the smaller ionic radius of Ca compared to the one of Sr ($r_{\rm Ca} = 1.18\,$\AA \, and $r_{\rm Sr} = 1.31\,$\AA) leads in Ca$_2$RuO$_4$ to substantial deviations from this configuration. Actually, with respect to the ideal tetragonal structure ($a$=$b$, $c$), the RuO$_6$ octahedra show alternating rotations about the direction, say $z$, of the apical Ru-O$_2$ bond, together with tilts of $z$ with respect to the $a$-$b$ plane initially containing the Ru-O$_1$ bonds ($x$ and $y$ axes), and distortions making $x$ and $y$ slightly different. As a result, the unit cell of Ca$_2$RuO$_4$ is orthorhombic and contains four Ru atoms (see Fig.~\ref{crystal}), the space group being the $Pbca$ one. In this configuration, Ca$_2$RuO$_4$ is a Mott insulator~\cite{Nakatsuji97a,Nakatsuji97b} with a gap estimated to be about 0.2 eV. An insulator-to-metal transition occurs at $T_{\rm IM}\simeq 360\,$K and is accompanied by a structural transition from an orthorombic insulating phase with a short $c$-axis length of the RuO$_6$ octahedra ($S$-$Pbca$), to a quasi-tetragonal metallic one with long $c$-axis length ($L$-$Pbca$)~\cite{Alexander99}. Clear evidence of this transition is given by the temperature dependence of the resistivity~\cite{Alexander99} and the $c$-axis lattice constant~\cite{Friedt01}, shown in the upper and the middle panel of Fig.~\ref{rho-c_axisT}, respectively. 
The $a$ and $b$-axes lattice constants also vary in the transition from the metallic to the insulating phase; in particular the $a$ axis contracts by 1.5$\,\%$ and the $b$ axis expands by 3$\,\%$ on cooling over an interval of 250 K (lower panel of Fig.~\ref{rho-c_axisT}). The combined effect of these thermally-induced deformations is to induce in the system an increasingly strong orthorhombic distortion that contracts the volume by 1.3$\,\%$ as $T$ is lowered from 400 K to 70 K~\cite{Alexander99,Cao97,Braden98,Cao00,Cirillo19}.  
The structural transition is so dramatic to break single-crystal samples into pieces when passing from the $L$-$Pbca$ structure to the $S$-$Pbca$ one~\cite{Nakamura2013}.

\begin{figure}[!bth]
\begin{center}
\includegraphics[width=0.8\textwidth]{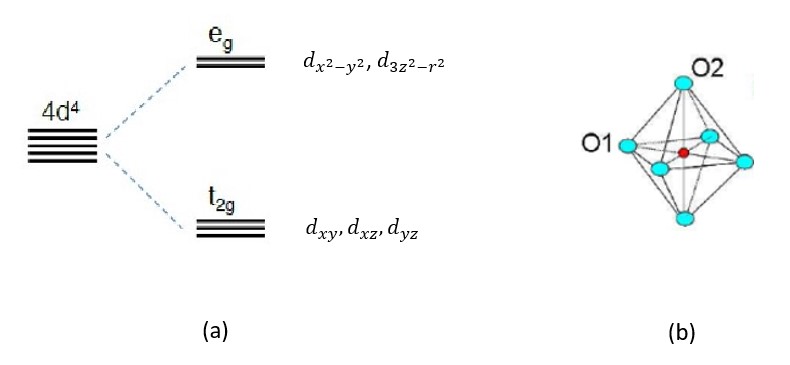}
\end{center}
\caption{(a) Energy levels of the Ru $4d$ orbitals as splitted by the Jahn-Teller effect; (b) RuO$_6$ fundamental crystal unit.}
\label{levels}
\end{figure}

In the RuO$_6$ octahedral structure, the vertices of the octahedron are occupied by six O atoms, with the Ru atom located at its center. Such a type of Ru-O coordination, according to the crystal field effect, splits the $d$ levels of the ruthenium atom in two groups:
the doublet $e_g$, higher in energy and including $d_{x^2-y^2}$ and $d_{3z^2-r^2}$ orbitals, and the triplet  $t_{2g}$, lower in energy and including $d_{xy}$, $d_{yz}$ and $d_{xz}$ ones (see Fig.~\ref{levels}\,a). The two orbitals of the $e_g$ doublet lie higher in energy because their lobes point exactly in the direction of the $p$ orbitals of O.
As for the lower three $t_{2g}$ levels, their degree of degeneracy is determined by the distances from the Ru atom of the in-plane O$_1$ oxygens and the apical O$_2$ oxygens (see Fig.~\ref{levels}\,b), denoted above by $z$ and $x$ and $y$, respectively,

A perfect octahedron ($z = x = y$) leads to an exact degeneracy, whereas 
denoting by $\overline x$ the average between $x$ and $y$, the smaller is the ratio $\overline x/z$, i.e. the more elongated in the $z$ direction the RuO$_6$ octahedron is, the higher in energy the $d_{xy}$ orbital is relative to the $d_{yz}$-$d_{xz}$ doublet~\cite{Cirillo19}.

\begin{figure}[!bth]
\begin{center}
\includegraphics[width=0.8\textwidth]{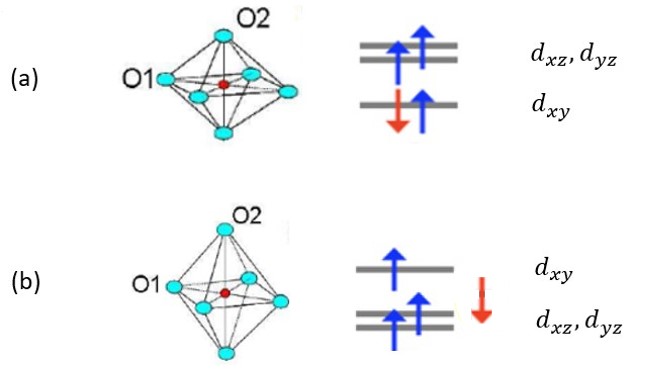}
\end{center}
\caption{(a) RuO$_6$ unit and $t_{2g}$ levels occupation in the insulating and in the metallic phase ((a) and (b), respectively). Blue and red arrows are used to distinguish spin-up and spin-down electrons.}
\label{levelsCF}
\end{figure}

It is well established that the low-energy properties of Ca$_2$RuO$_4$ are determined by the four $4d$ electrons of the Ru$^{4+}$ ions. The order in energy of the $t_{2g}$ levels, which can of course accommodate up to six electrons, is therefore fundamental in determining how these electrons occupy the levels, thus establishing the transport properties of the resulting state.
In the insulating short $c$-axis state ($z/\overline x < 1$), the orbital $d_{xy}$ is lower in energy than the $d_{yz}$-$d_{xz}$ doublet, with a crystal field gap sufficiently large that the electrons prefer to arrange in pairs in the $d_{xy}$ level, although the local Coulomb repulsion would hinder this arrangement (see Fig.~\ref{levelsCF}\,(a)). The remaining two electrons occupy the $d_{yz}$-$d_{xz}$ doublet with parallel spins, in accordance with Hund’s rule, and this configuration leads to an insulating antiferromagnetic state at sufficiently low temperatures. 
When higher temperatures are considered, the ratio $z/\overline x < 1$ approaches 1 from below, and the three levels tend to become almost degenerate. Nonetheless, strong correlations still make the system behave as an insulator by splitting the $d_{xz}$ and $d_{yz}$ levels and originating from them lower, completely filled, and upper, completely empty, Mott-Hubbard bands~\cite{Anisimov02,Gorelov10}. A further increase of the temperature makes the ratio $z/\overline x$ becomes larger than 1, enough to push the $d_{xz}$-$d_{yz}$ doublet below the $d_{xy}$ level. In this way, the system becomes metallic, with three electrons filling the lowest levels according to the Hund’s rule, and the fourth electron becoming free to move within the lattice (Fig.~\ref{levelsCF}\,(b)).

When higher temperatures are considered, the ratio $z/\overline x < 1$ approaches 1 from below, and the three levels tend to become almost degenerate. Nonetheless, strong correlations still make the system behave as an insulator by splitting the $d_{xz}$ and $d_{yz}$ levels, leading to lower, completely filled, and upper, completely empty Mott-Hubbard bands. A further increase in temperature makes the ratio $z/\overline x$ exceed 1, which is sufficient to push the $d_{xz}$-$d_{yz}$ doublet below the $d_{xy}$ level. In this way, the system becomes metallic, with three electrons filling the lowest levels according to Hund’s rule, and the fourth electron becoming free to move within the lattice (Fig.~\ref{levelsCF}(b)).

\begin{figure}[!bth]
\begin{center}
\includegraphics[width=0.6\textwidth]{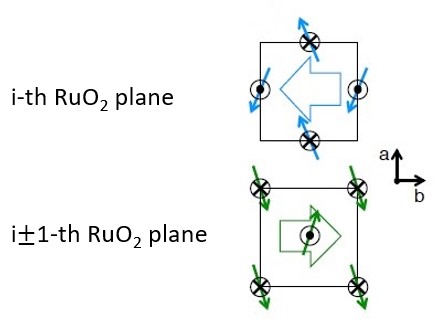}
\end{center}
\caption{Spin configuration in the A-centered mode antiferromagnetic phase developing below $T_N=110\,$K (adapted with permission from Ref.\cite{Fukazawa01}, copyrighted by the Physical Society of Japan).}
\label{AFM}
\end{figure}

A further relevant element affecting the electronic properties of Ca$_2$RuO$_4$ is the spin-orbit coupling. As is well known, compared to the Coulomb repulsion $U$ and to the crystal field potential $\Delta$, this interaction is negligible in systems with partially filled $3d$ shells, is strong in $5d$ systems and comparable with $U$ and $\Delta$ in $4d$ ones. For this reason, the competition of spin-orbit coupling and crystal-field splitting in a system such as Ca$_2$RuO$_4$ that contains ruthenium, has been discussed controversially over many years~\cite{Mizokawa01,Fatuzzo15,Lotze21,Vergara22}.
A recent study~\cite{Vergara22} seems to indicate that Ca$_2$RuO$_4$ is firmly
rooted in the range of dominant crystal field, but with the spin-orbit coupling still playing an important role.

\section{Altermagnetic phase}

It is well known that magnetism occupies a central role in condensed matter physics. 
With regard to the two ordered phases by far most studied in this field, ferromagnetism and antiferromagnetism, one has in the former case a spin polarization that breaks the time-reversal symmetry and leads to spin-splitting of the band structure, and in the latter a net zero magnetization with the band structure showing no nonrelativistic spin-splitting. 

Very recently, a new type of magnetic phase has been discovered, named altermagnetism. In altermagnets, the magnetization in the real space is zero, due to antiparallel orientation of spins, like in antiferromagnets, but the band structure presents non-relativistic spin-splitting, like in ferromagnets~\cite{Smejkal22,Smejkal22b}. In this sense, altermagnets are antiferromagnets that do not exhibit Kramers degeneracy. In these systems, the electronic charge of the spin-up (down) atoms should be mapped into the charge of the spin-down (up) atoms without translations, inversion, or their combinations, but only with roto-translations or mirror reflections. In this context, the role of the non-magnetic atoms and the way they influence magnetic symmetries is crucial. Even without the presence of a ferromagnetic order in real space, the altermagnets present anomalous Hall effect, due to the spin-splitting of the bands~\cite{Smejkal20,Betancourt23}. 
Magnetic space groups are classified into four types based on their relationship to the parent crystallographic space group~\cite{Bradley10}. Recent studies highlight that altermagnetism is present in type-I and type-III magnetic space groups~\cite{Guo23}.

Various transition metal oxides~\cite{Guo23,Chen23} and rare-earth compounds~\cite{Cuono23Eu} have been identified as exhibiting this phenomenon, included Ca$_2$RuO$_4$~\cite{Cuono23,Sattigeri23}.
Previous studies have shown that Ca$_2$RuO$_4$ is characterized by canted antiferromagnetism below the N\'eel temperature $T_N=110\,$K~\cite{Cao97}, originating from oxygen-mediated superexchange coupling between Ru $4d$-electrons. The transition at $T_N$ is signalled by cusps in the temperature dependence of the magnetic susceptibility, measured on single crystals with the field $H$ both parallel and perpendicular to the $c$ axis~\cite{Cao97}. The values of $M/H$ ($M$ is the magnetization) observed with $H\parallel c$ are larger than in the case $H\perp c$, this indicating that the easy axis for the antiferromagnetic order lies in the $a$-$b$ plane. Evidence for antiferromagnetic ordering below 110 K is also provided by measurements of the field dependence of the magnetization performed at fixed temperature. When this is done at $T=105\,$K, $M(H)$ exhibits an hysteretic transition below $H\simeq 9\,$T~\cite{Cao97} which is absent at a lower temperature. Moreover, neutron diffraction studies~\cite{Braden98,Fukazawa01} have shown that the in-plane alignment of the canted spins gives rise to magnetic moments that can be either parallel or antiparallel to the crystallographic $b$ axis, with opposite directions in two consecutive planes (A-centered mode, see Fig.~\ref{AFM}). A minor antiferromagnetic phase (B-centered mode) develops between approximately 110 K and 150 K, with the spins arranged in such a way to give moments in two consecutive planes pointing in the same direction, rather in the opposite one. Resonant elastic x-ray scattering measurements showed that, even if in the ground state the magnetic moment is oriented along the $b$ axis, a small canting along the $c$ axis is also present \cite{Porter18}. No signal of a magnetic moment along the $a$ axis has been found \cite{Porter18}.

Altermagnetism in Ca$_2$RuO$_4$ has been studied by ab initio methods~\cite{Cuono23,Sattigeri23} briefly described below. We recall that similar approaches, often based on a combination of Local Density Approximation (LDA) and Dynamical Mean-Field Theory (DMFT), have been widely used for the investigation of the electronic and magnetic properties of Ca$_2$RuO$_4$, with a special attention to the connection between the metal-insulator transition which characterizes the behavior of this compound and the structural transition which accompanies the phase change~\cite{Gorelov10,Zhang17}. 
The computational details adopted in the investigation of this phenomenon and the main results correspondingly obtained are described below in two separate subsections.

\subsection{Computational details}

We performed all calculations without taking relativistic effects into account, employing the Vienna ab initio simulation package (VASP)~\cite{Kresse93,Kresse96,Kresse96b}. LDA well reproduces the electronic properties in the paramagnetic phase, but in the $S$-$Pbca$ phase, which is antiferromagnetic (AFM) and insulating, it still gives a metallic ground state. To take into account the correlations in the Ru-4$d$ states, a Coulomb repulsion should be considered in the calculations. To this purpose, we have used the generalized gradient approximation (GGA) of Perdew, Burke and Ernzerhof~\cite{Perdew96}, focusing on the analysis of the $S$-$Pbca$ phase, characterized by space group number 61. A plane-wave energy cut-off of 480 eV was utilized, ensuring total energy minimization to less than 1 $\times$ 10$^{-5}$ eV. We performed the computation employing 11 $\times$ 11 $\times$ 4 $k$-points, considering the experimental lattice constant values $a$ = 5.3945 {\AA}, $b$ = 5.5999 {\AA}, $c$ = 11.7653 {\AA}~\cite{Friedt01}. A Coulomb repulsion $U$ = 3 eV was employed on the Ru-4$d$ orbitals~\cite{Autieri16}, with a Hund's coupling $J_{\rm H}$ = 0.15$\;U$~\cite{Vaugier12}. Notably, previous studies on bulk ruthenates employed a Coulomb repulsion value smaller than 1 eV~\cite{Etz12,Rondinelli08}. Moreover, in the insulating phase, reduced screening led to a higher Coulomb repulsion value. Within this framework, the magnetic moment per Ru atom was determined to be 1.4 $\mu$B, consistently with the typical magnetic configuration in Ca$_2$RuO$_4$.

\subsection{Orbital-selective altermagnetism}

\begin{figure}[t!]
\centering
\includegraphics[width=8cm,angle=0]{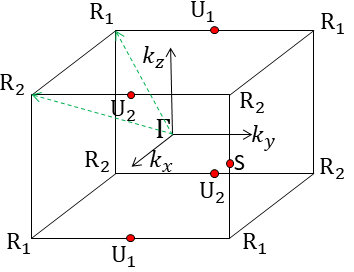}
\caption{Symmetries of the irreducible Brillouin zone for the orthorhombic Ca$_2$RuO$_4$. With the subscripts 1 and 2 we indicate the k-points that have opposite non-relativistic spin-splitting along the path leading to the $\Gamma$ point. The dashed arrows indicate one possible path where the band structure shows altermagnetic spin-splitting. }\label{Brillouin zone}
\end{figure}

Altermagnetism in Ca$_2$RuO$_4$ has an orbital-selective origin, where the d$_{xy}$ bands are not splitted, while the d$_{xz}$/d$_{yz}$ band structure subsector presents non-relativistic spin-splitting~\cite{Cuono23}. Due to the splitting of the bands, the altermagnetic systems can show anomalous Hall effect once the relativistic effects are considered. 

\begin{figure}[t!]
\centering
\includegraphics[width=5cm,angle=270]{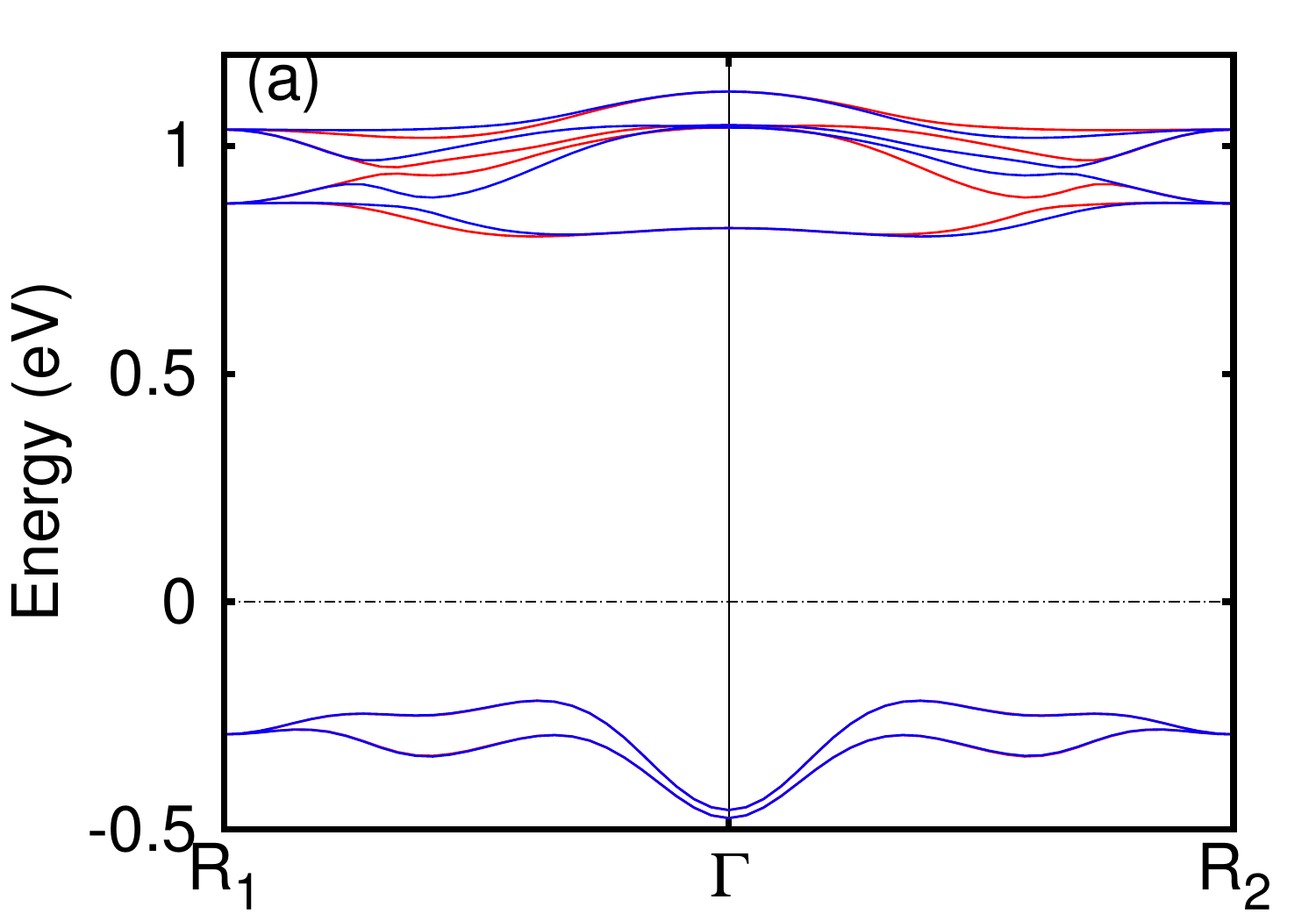}
\includegraphics[width=5cm,angle=270]{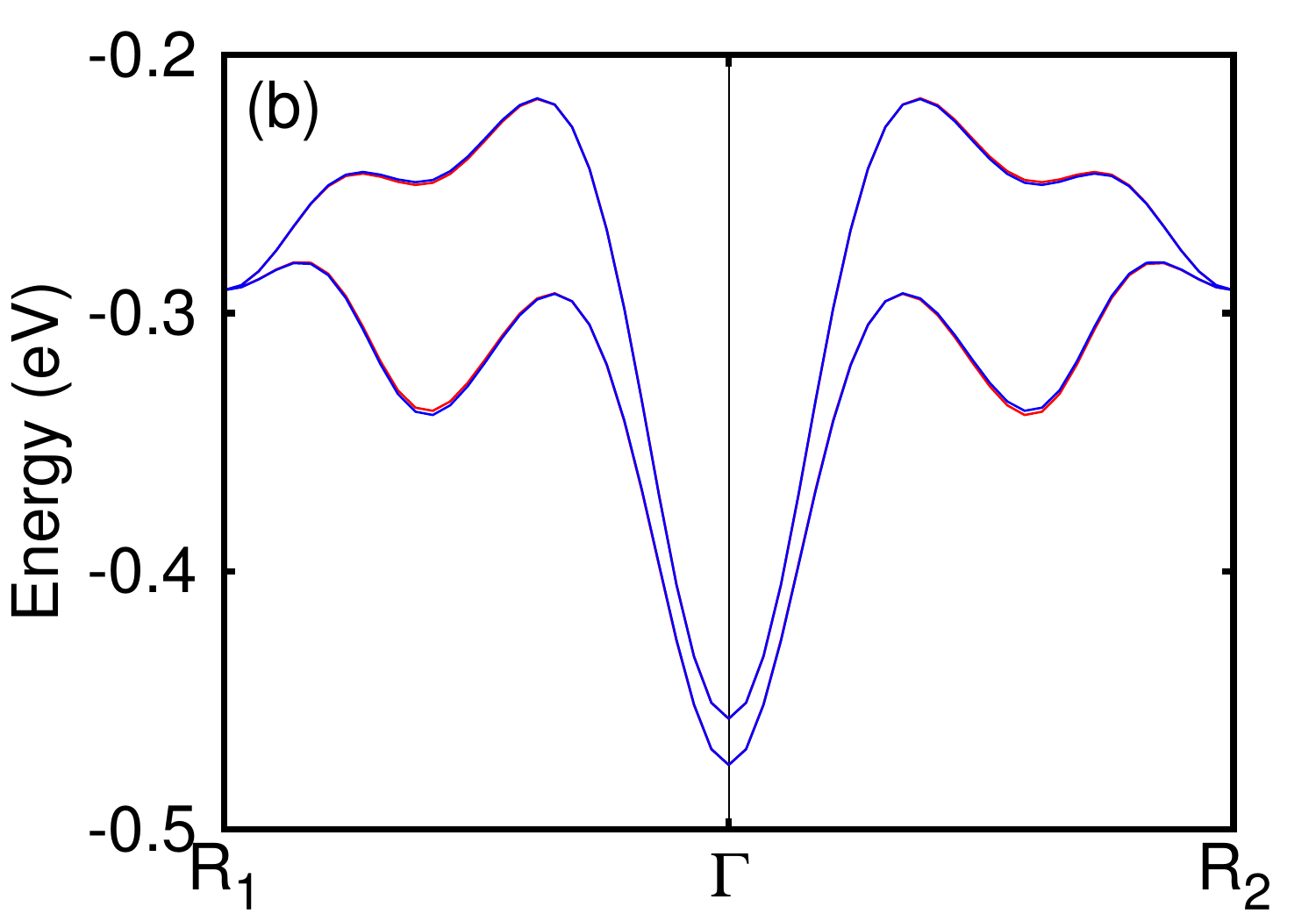}
\includegraphics[width=5cm,angle=270]{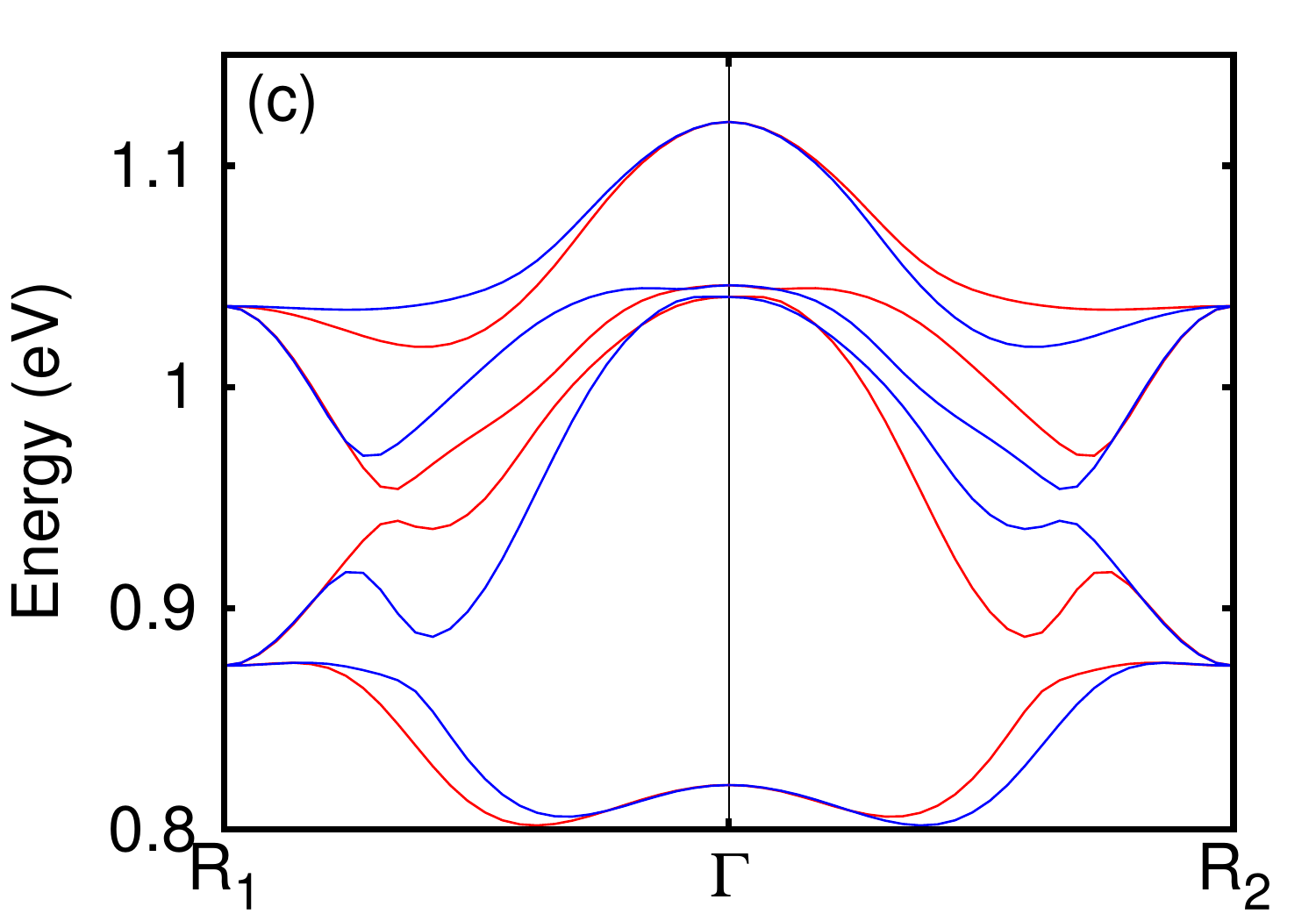}
\includegraphics[width=5cm,angle=270]{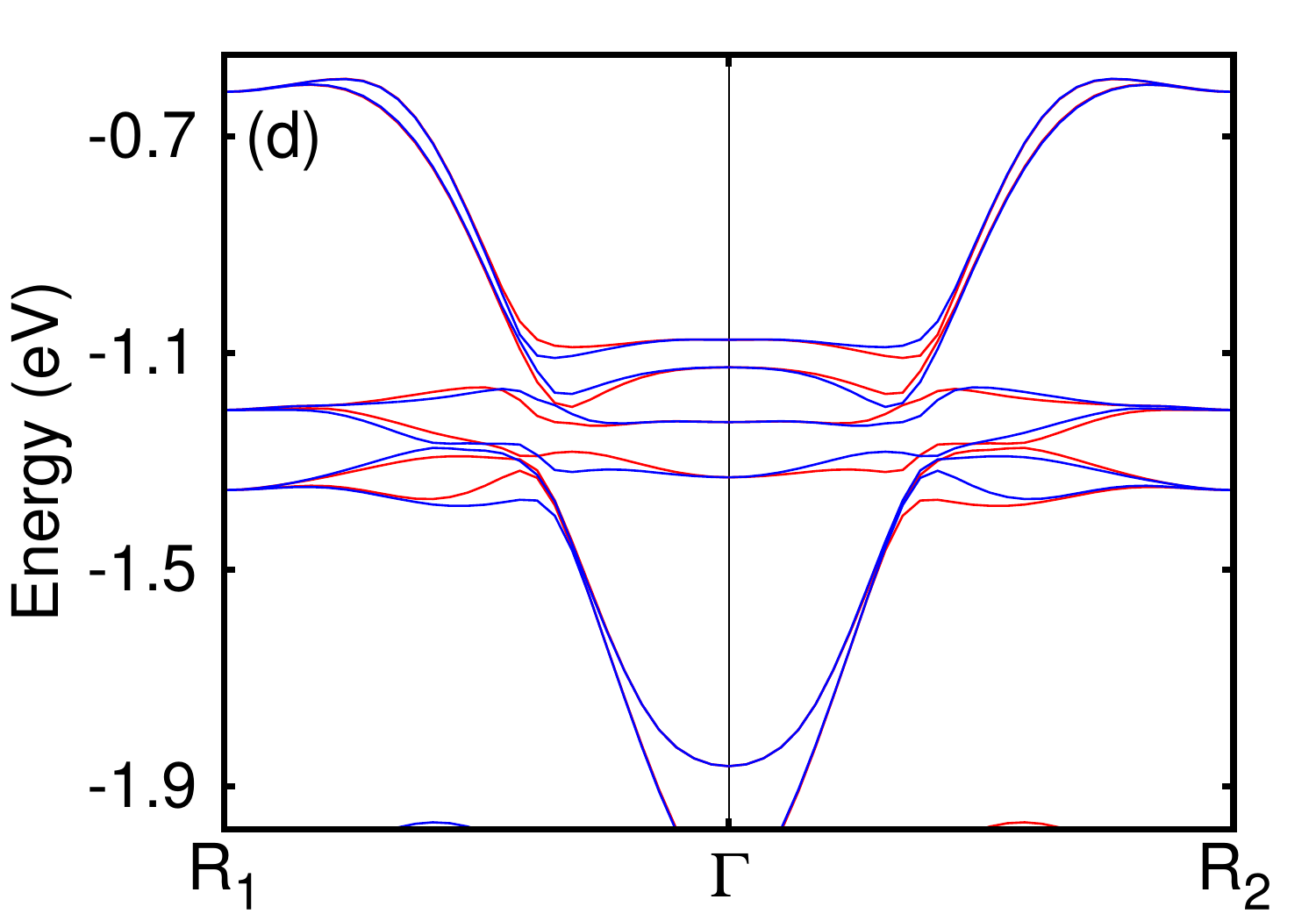}
\caption{(a) Band structure of Ca$_2$RuO$_4$ along the high-symmetry path R$_1$-$\Gamma$-R$_2$ plotted in the energy range from -0.5 to 1.2 eV. 
(b) Magnification of the band structure shown in panel (a) in the energy range between -0.2 to -0.5 eV.
(c) Same as in panel (b) in the energy range between 0.8 and 1.15 eV.
(d) Magnification of the band structure along the R$_1$-$\Gamma$-R$_2$ k-path with energy between -1.98 and -0.55 eV.
Blue and red lines denote spin-up and spin-down channels, respectively.
The Fermi level is set to zero energy.}\label{BS_CAO_altermagnetism}
\end{figure}

\begin{figure}[t!]
\centering
\includegraphics[width=6cm,angle=270]{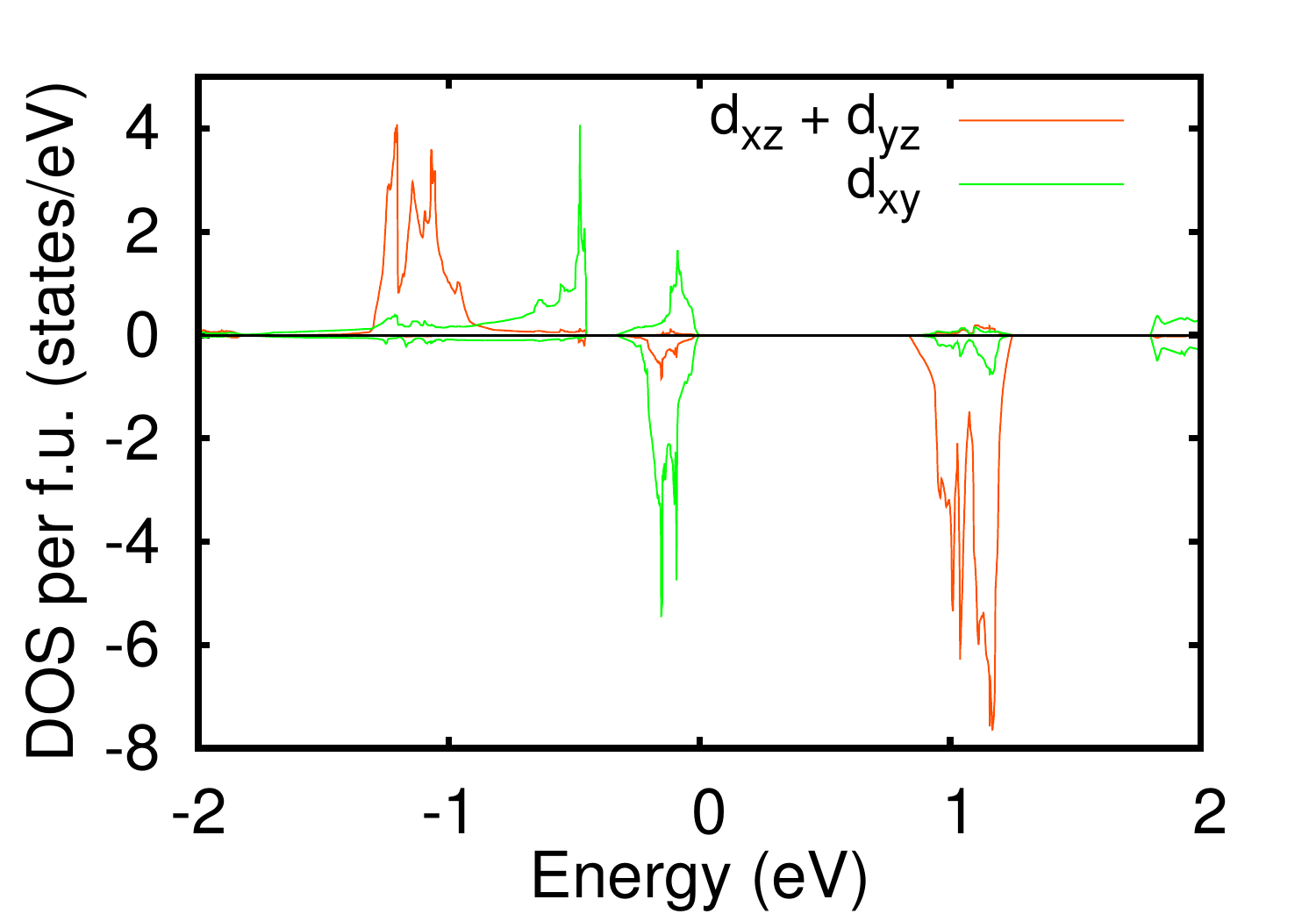}
\caption{{Local densities of states for the t$_{2g}$ orbitals of the Ru atoms.
Those associated with the d$_{xz}$/d$_{yz}$ and with the d$_{xy}$ orbitals are plotted in red and green, respectively. 
We report the LDOSs for the spin-up subsector with positive values, and those for the spin-down subsector with negative values. The Fermi level is set to zero energy.}}\label{DOS_t2g}
\end{figure}

\begin{figure}[t!]
\centering
\includegraphics[width=6cm,angle=270]{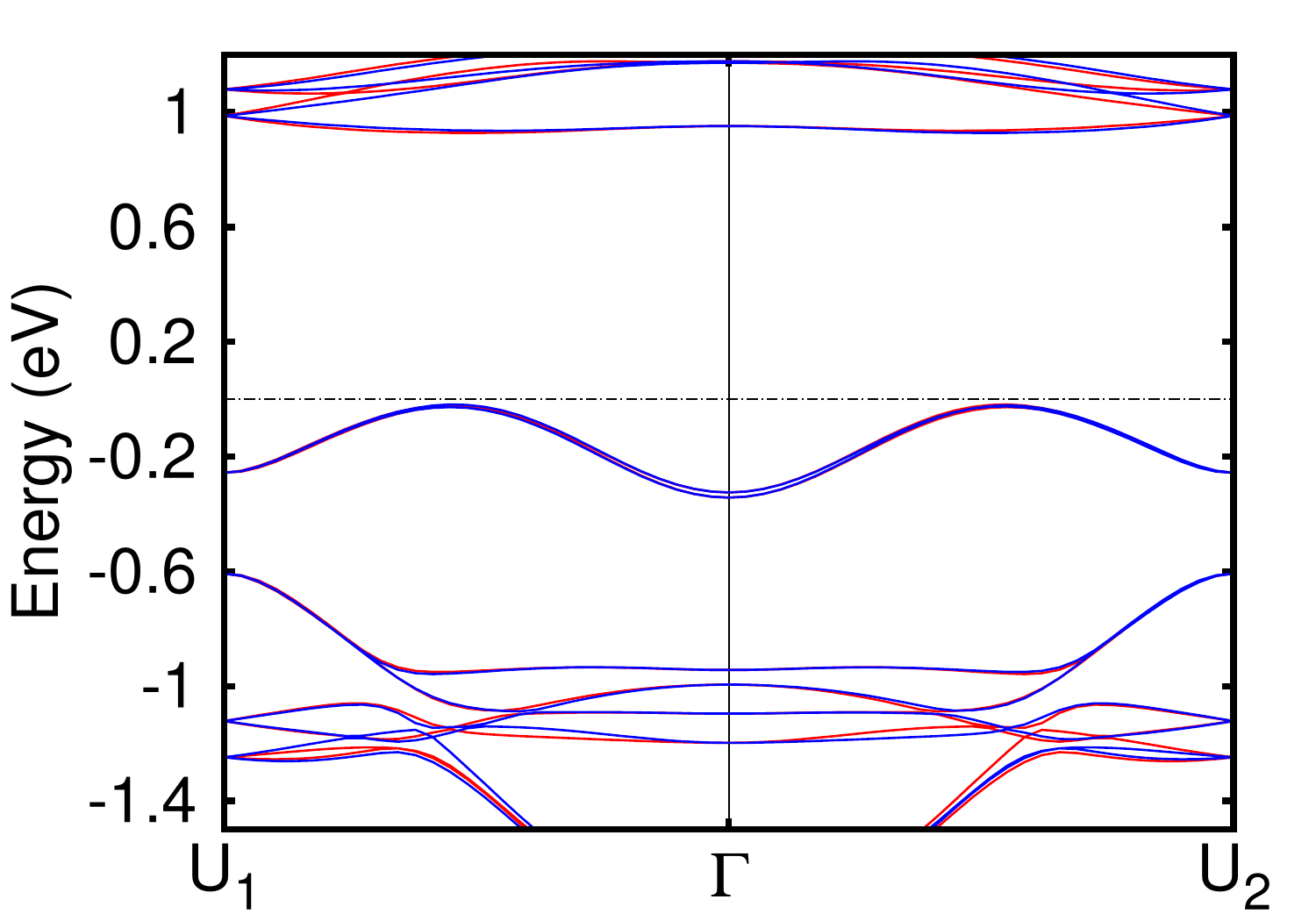}
\caption{{Band structure of Ca$_2$RuO$_4$ along the high-symmetry path U$_1$-$\Gamma$-U$_2$ plotted in the energy range from -1.5 to 1.2 eV.
The Fermi level is set to zero energy.}}\label{BS_CAO_U_G}
\end{figure}

\begin{figure}[t!]
\centering
\includegraphics[width=6cm,angle=270]{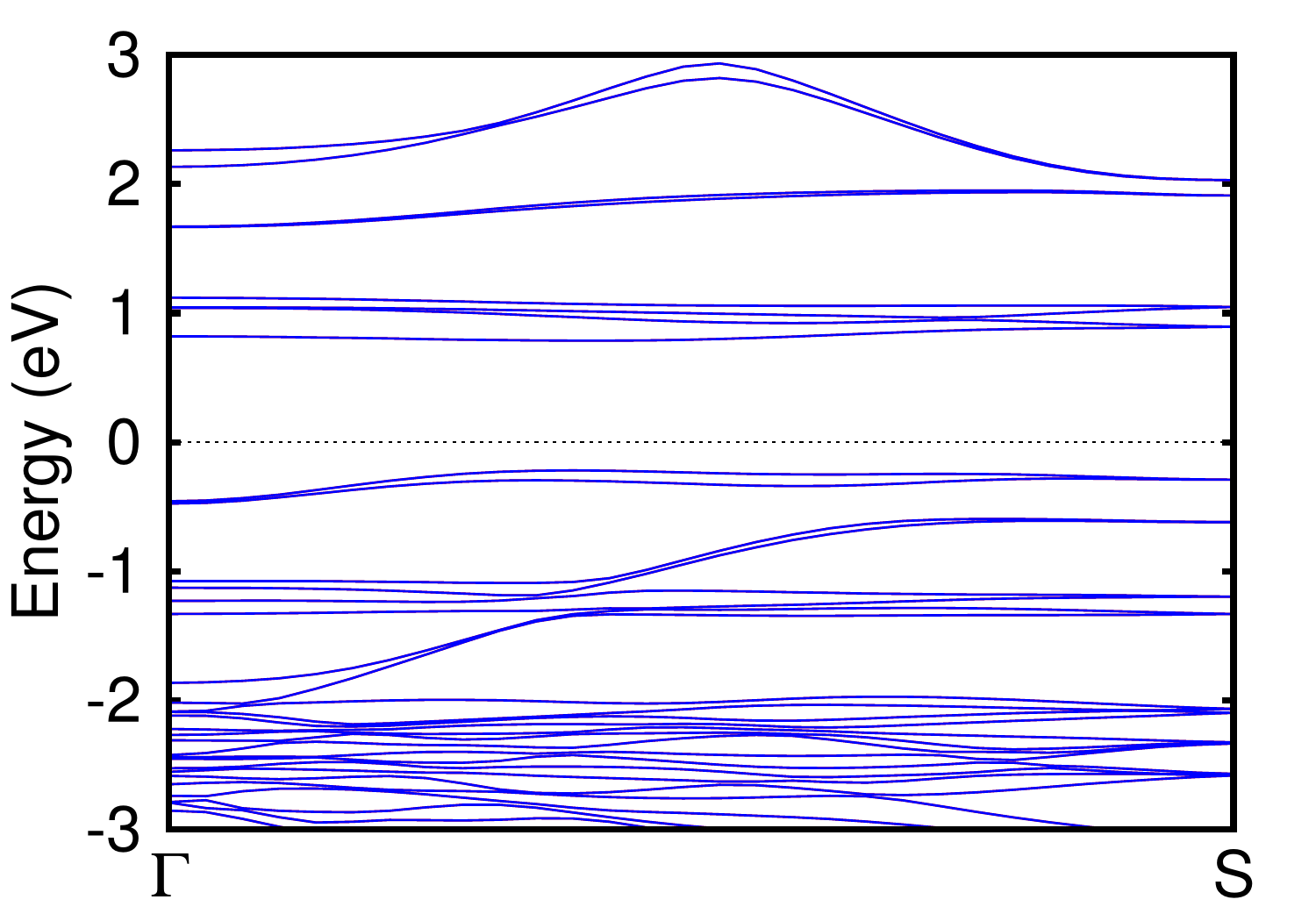}
\caption{Band structure of Ca$_2$RuO$_4$ along the high-symmetry line $\Gamma$-S plotted in the energy range from -3 to 3 eV. The band structure does not present non-relativistic spin-splitting along this line.
The Fermi level is set to zero energy.}\label{BS_CAO_G_S}
\end{figure}

The altermagnetic spin-splitting is strongly momentum-dependent, it is present when we consider the antiferromagnetic configuration of the system and only along some lines of the Brillouin zone.
In Fig.~\ref{Brillouin zone}, we report the orthorombic Brillouin zone and the points that have opposite altermagnetic spin-splitting in the AFM phase of Ca$_2$RuO$_4$.
We have shown that the band structure presents non-relativistic spin-splitting along the lines R$_1$-$\Gamma$-R$_2$ and U$_1$-$\Gamma$-U$_2$, which are the points reported in Fig. \ref{Brillouin zone}. In Fig.~\ref{BS_CAO_altermagnetism} we report different zooms of the band structure along the path R$_1$-$\Gamma$-R$_2$, denoting with blue and red lines the contributions associated with the spin up and down channel, respectively.
We can see that while some bands do not split, others on the contrary do, with the spin-splitting along $\Gamma$-R$_2$ being opposite with respect to the splitting along $\Gamma$-R$_1$.
The orbital character of the band structure has been investigated in literature. 
There are two sets each consisting of four d$_{xz}$/d$_{yz}$ bands around -$1\;$eV and +$1\;$eV, which are less dispersive, and two sets each consisting of two more dispersive d$_{xy}$ bands, developing between -0.5 eV to -0.2 eV and between -2 eV and -0.6 eV,  respectively. The latter hybridizes with the majority d$_{xz}$/d$_{yz}$ bands around -1 eV, as one can see from Fig.~\ref{BS_CAO_altermagnetism}(d). The contributions of the t$_{2g}$ orbitals is evident from the local density of states (LDOS) reported in Fig.~\ref{DOS_t2g}. The character of the related bands close to the Fermi level as described in Section 3 is confirmed, namely the bands close to -1 eV and 1 eV are mainly d$_{xz}$/d$_{yz}$, while the dispersive bands between -0.5 eV to -0.2 eV and between -2 eV and -0.6 eV are mainly d$_{xy}$. The d$_{xy}$ bands are not splitted, as we can clearly see in Fig. \ref{BS_CAO_altermagnetism}(b), while the d$_{xz}$/d$_{yz}$ present an altermagnetic spin-splitting (Fig.~\ref{BS_CAO_altermagnetism}(c)). This is a very interesting case of orbital-selective altermagnetism, where non-relativistic spin-splitting is present or absent depending on the orbital character of the bands. While the d$_{xy}$ orbitals are quasi two-dimensional because of negligible hopping amplitudes between the layers, this is not true for the d$_{xz}$/d$_{yz}$ orbitals. It has already been shown in the literature that bidimensional systems tend not to be altermagnetic for symmetry reasons~\cite{Cuono23,Sattigeri23}. Therefore, here the quasi-two dimensional d$_{xy}$ orbitals are not affected by altermagnetism. We have checked the presence of altermagnetic spin-splitting along other lines of the Brillouin zone, finding that it is also present along the $\Gamma$-U line. The band structure along the path U$_1$-$\Gamma$-U$_2$ is plotted in Fig.~\ref{BS_CAO_U_G}. We can clearly see the presence of orbital-selective altermagnetism even along these lines. As along the path R$_1$-$\Gamma$-R$_2$, the bands show opposite spin-splitting along $\Gamma$-U$_1$ and $\Gamma$-U$_2$, and the non-relativistic spin-splitting is present for the d$_{xz}$/d$_{yz}$ orbitals, while it is absent in the d$_{xy}$ sector. While splitting is maximum along $\Gamma$-R and $\Gamma$-U lines, it also survives in all the Brillouin zone except in the high-symmetry directions where one of the k-components is zero or on the zone boundaries. We point out that the non relativistic spin-splitting is only present along some lines of the Brillouin zone. As an example, we have that it is not present along the $\Gamma$-S line, as shown in Fig.~\ref{BS_CAO_G_S}.

In summary, Ca$_2$RuO$_4$ is an altermagnetic system that presents non-relativistic spin-splitting only for some bands, thus showing a very peculiar orbital-selective form of  altermagnetism. It is important to underline that altermagnetic systems can be used in many applications, going from spintronics~\cite{Shao21} to  spincaloritronics~\cite{Zhou24}, and they can  contribute to highly efficient spin-current generation~\cite{Hernandez21}, produce giant and tunneling
magnetoresistance~\cite{Smejkal22magneto} and can be useful in Josephson junctions~\cite{Ouassou23}.

\section{Negative thermal expansion}

\begin{figure}[t!]
\centering
\includegraphics[width=0.7\textwidth,angle=0]{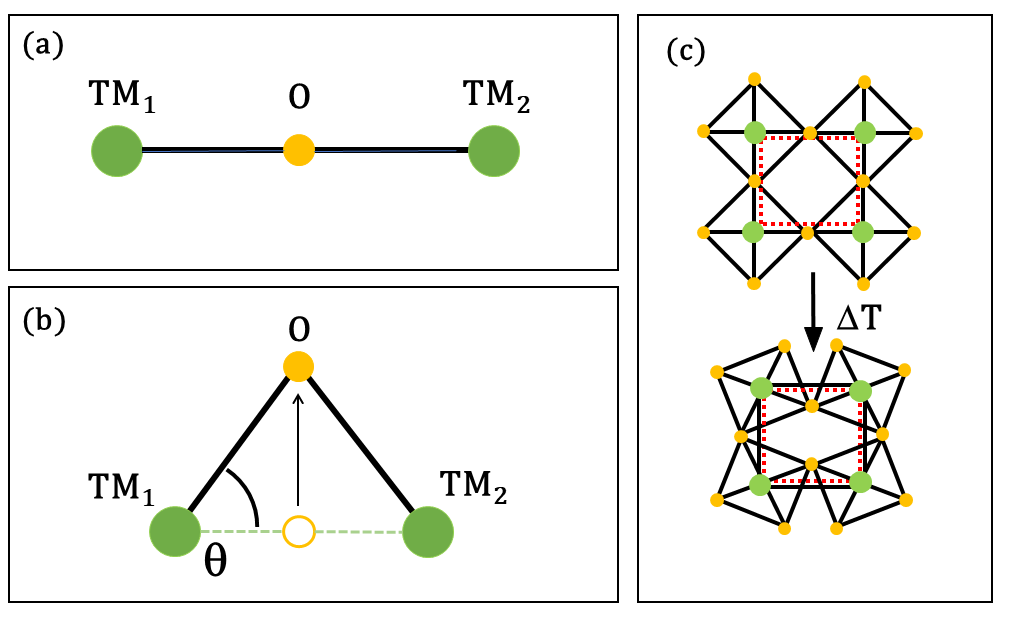}
\caption{{TM$_1$-O-TM$_2$ bond for
undistorted configuration, i. e. for vanishing bond angle $\theta$ (panel (a)), and for distorted configuration with $\theta\ne 0$ (panel (b)). In panel (c) it is shown that a thermal gradient, which causes an increase in the bond angle, results in a reduction of the distance between the TM ions, thereby decreasing the unit cell area.}}\label{bond}
\end{figure}

In insulating phases marked by electron correlation and multi-orbital configurations, the Coulomb interaction obstructs charge mobility, with kinetic energy reliant on virtual charge transfer and spin-orbital exchanges. Unlike single-orbital models, forecasting configurations in multi-orbital states is arduous and necessitates comprehension of how orbital-dependent magnetic correlations impact transition metal-oxygen-transition metal (TM-O-TM) distortions, thereby establishing the conditions for NTE emergence. The phenomenon of NTE encompasses the decrease in lattice parameters upon heating or conversely, the expansion during thermal cooling. Although not widely observed, NTE bears significant implications for various fields including electronics, optics, and thermal engine or medical product design. Beyond structural shifts, the complexity of NTE origin resides in the intricate interplay among electron, spin, and orbital degrees of freedom. For instance, in certain Zn-based materials, NTE predominantly originates from oxygen vibrational modes and anharmonicity associated with spin canting and spin-lattice coupling~\cite{Hua16,Martineck68,Mary96,Ernst98}. Consequently, while alterations in lattice parameters directly incite NTE, a fundamental inquiry arises regarding whether electronic correlations and ensuing spin-orbital couplings synergize or counteract the lattice's propensity to demonstrate NTE. In perovskite oxides featuring transition metals within octahedral oxygen cages, alterations in unit cell size are tied to octahedral rotations, and thus to the value of TM-O-TM bond angles. The undistorted or distorted nature of these bonds, delineated by zero or non-zero bond angles, correlates with unit cell volume. The TM-O-TM bond angle plays a pivotal role in governing NTE, influencing electron connectivity within the lattice via orbital hybridization between transition metal and oxygen atoms. Specifically, since the distance between TM ions relates to the TM$_1$-O-TM$_2$ bond angle, itself contingent upon rotation of TM-O octahedra around a given crystallographic axis, deviation from 180$^{\circ}$ of the TM$_1$-O-TM$_2$ bond angle yields a reduction in lattice parameters, {as illustrated by the comparison between the undistorted (Fig.~\ref{bond}(a)) and distorted (Fig.\ref{bond}(b)) cases. In Fig.~\ref{bond}(c), we schematically represent how a variation of the bond angle in a square geometry of the TM-O lattice can lead to a modification of the unit cell area. Specifically, we show how a thermal gradient that induces an increase in the bond angle would lead to a reduction in the distance between the TM ions and, in turn, of the unit cell area. When this picture is referred to real three-dimensional systems, the described mechanism leads to an increase of the unit cell volume, thus giving rise to NTE.}

In materials featuring Mott physics and intricate magnetic ordering, the interplay of electronic degrees of freedom becomes crucial, particularly in configurations characterized by significant spin-orbital entanglement. For instance, the observed negative thermal expansion (NTE) in insulating Ca$_2$RuO$_4$ is believed to stem from electronic correlations, where spins and orbitals exhibit independent dynamics but are coupled through lattice distortions. Substituting a small fraction of Ru with Cr induces a notable 1$\%$ volume reduction due to the collapse of orbital and magnetic ordering~\cite{Qi10}. Similar NTE effects are observed in compounds like Ca$_2$Ru$_{1-x}$TM$_x$O$_4$ (TM = Mn, Fe, or Cu), where the substitution correlates with metal-insulator and magnetic ordering transition temperatures~\cite{Qi12,Tak17}. NTE is also found in Ruddlesden-Popper calcium titanates Ca$_{n+1}$Ti$_n$O$_{3n+1}$ family ($n=1,2,3$)~\cite{Koo21}.
Therefore, theoretical investigations into NTE must consider three primary factors: the bond angle in $d$-$p$ hybridization processes, the symmetry properties of the compound, and Hund's rule coupling. Concerning the bond angle, anisotropic orbital correlations, stronger at lower temperatures with robust spin-orbital correlations, tend to drive bond direction towards lesser distortion. Conversely, thermal fluctuations weaken spin-orbital correlations, promoting more distorted bonds. Thermal disorder also induces non-collinear magnetic correlations, energetically favoring distorted bonds and resulting in NTE effects.
In this context, the existence of low point group symmetry conditions plays a crucial role. Breaking C$_4$ rotation symmetry, especially in the presence of strong Coulomb interaction, can transform positive thermal expansion into NTE behavior, particularly when tracking the TM-O-TM bond angle. This phenomenon is exemplified by magnetic patterns that break C$_4$ rotation symmetry. The study predicts similar outcomes for NTE when considering orbital or electronic patterns that reduce point group symmetry, especially in thin films under strain or exposed to an electric field.
Lastly, Hund's coupling emerges as another critical element in NTE effects. Indeed, the increase of Hund's coupling amplitude drives a  transition from positive to negative thermal effects, underlining its significance in multi-orbital materials. The analysis suggests that a large Hund's coupling may promote NTE occurrence in doped Hund's metals with strong correlations.

\subsection{The model}
The determination of the energetically most favorable value of the TM-O-TM angle is a complex issue influenced by multiple factors, including charge distribution, spin states, and orbital interactions. Specifically, we investigate in this context the role played by the Coulomb interaction at the TM site, in connection with spin-orbital correlations or other electronic parameters associated with tetragonal distortions.
Our analysis involves the exact solution of a model Hamiltonian for the TM$_1$-O-TM$_2$ cluster, featuring two TM ions at its ends~\cite{Brzezicki23}. Its expression is the following
\begin{equation}
H=H_{\rm{TM}}+H_{\rm{O}}+H_{{\rm TM}-{\rm O}} \; ,
\label{Ham_NTE}
\end{equation}
where $H_{\rm{TM}}$ includes the Coulomb interactions between $t_{2g}$ electrons, the tetragonal crystal field potential and the spin-orbit coupling~\cite{Cuoco06a,Cuoco06b,Forte10}; $H_{\rm{O}}$
contains the on-site energy terms for oxygen orbitals and, finally, $H_{{\rm TM}-{\rm O}}$ describes the TM-O hopping term. The explicit form of the above three terms is given in the Appendix.

The evolution with temperature of the phenomena determined by the microscopic interactions appearing in the above Hamiltonian will be studied using the Helmholtz free energy, given by
\begin{equation}
F_{el}(\theta)=-\frac{1}{\beta}\ln\left\{\sum_i\exp\left(-\beta E_i(\theta)\right)\right\}\, .
\end{equation}
Here $\beta=1/k_B T$ is the Boltzmann factor expressed in terms of the temperature $T$ and the Boltzmann constant $k_B$, while $E_i(\theta)$ are the eigenvalues of the Hamiltonian (\ref{Ham_NTE}), which depend on the bond angle $\theta$ and are evaluated by exact diagonalization.

\subsection{Discussion of the results}
\begin{figure}[t!]
\centering
\includegraphics[width=1\textwidth]{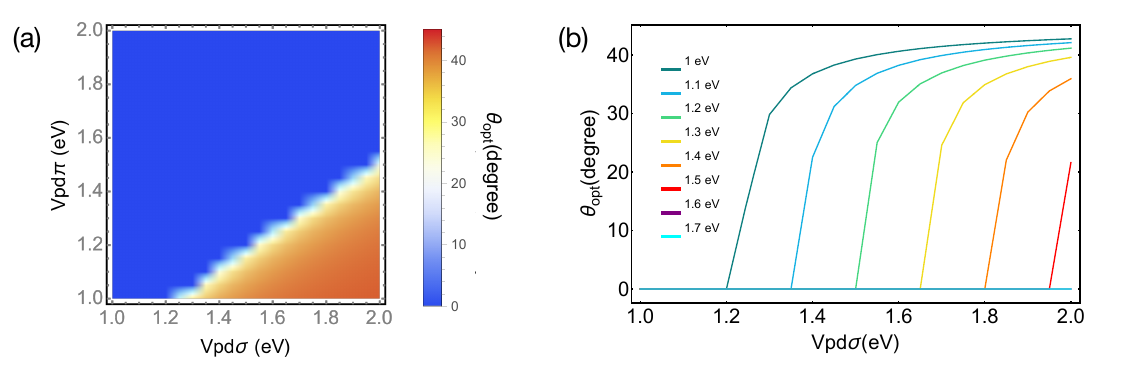}
\caption{(a) Zero-temperature contour map of the optimal TM($d^4$)$-$O$-$TM($d^4$) bond angle as a function of the $p$-$d$ hybridization parameters
$\{V_{pd\sigma}, V_{pd\pi}\}$ for $U=2.3$ eV, $J_{H}=0.5$ eV and $\varepsilon_{x,y,z}=-4.5$ eV. (b) Dependence of the optimal angle on $V_{pd\sigma}$ for fixed values of $V_{pd\pi}$ (the values of the other parameters are the same as in panel (a).}
\label{d3-d3}
\end{figure}

Hereafter, we present the outcomes of our numerical computations, with $U=2.3$ eV, $J_H=0.5$ eV, and an assumed value of $\varepsilon_{x,y,z}=-4.5$ eV for the oxygen orbitals. Here $U$ is the intra-site Coulomb interaction between electrons belonging to the same orbital, $J_H$ is the Hund's coupling, and $\varepsilon_{x,y,z}$ are the on-site energies for the oxygen $p_x$, $p_y$ and $p_z$ orbitals, respectively (see also the Appendix). Initially, we examine the system at zero temperature, varying the $p$-$d$ hybridization free parameters $V_{pd\sigma}$ and $V_{pd\pi}$. The extent of tetragonal distortions is quantified by the amplitude $\delta$, where $\delta=\varepsilon_{xy}-\varepsilon_{\gamma z}$ with $\varepsilon_{\gamma z}\equiv \varepsilon_{xz}=\varepsilon_{yz}$. Additionally, we explore the potential for orthorhombic splitting, denoted by $\delta_{\rm ort}$, in the ${xz,yz}$ orbitals, by assuming $\varepsilon_{yz}=\varepsilon_{\gamma z}+\delta_{\rm ort}$ and $\varepsilon_{xz}=\varepsilon_{\gamma z}-\delta_{\rm ort}$. In Fig.~\ref{d3-d3}, we depict the distribution of the optimal ground state bond angle $\theta_{\rm opt}$ for the TM($d^4$)-O-TM($d^4$) configuration. This figure illustrates the impact of $p$-$d$ covalency on the distortions observed in the TM-O-TM bond. Two distinct regimes of bond distortions, driven by electronic correlations, are discernible. Specifically, dominance of the $p$-$d$$\pi$ bonding amplitude ($V_{pd\pi}$) over the $\sigma$ bonding one ($V_{pd\sigma}$) leads to a preference for a small tilt in the bond angle, close to zero. Conversely, across the remaining phase diagram, a large bond angle of approximately $\pi/4$, resulting in all bonds being $\pi/2$, is favored. The transition between these regimes is abrupt and typically occurs when $V_{pd\sigma}$ is roughly equal to $V_{pd\pi}$. We also confirm that variations in other electronic parameters, such as $U$, $J_H$, etc., minimally impact the phase diagram.

\begin{figure}[t!]
\centering
\includegraphics[width=0.9\columnwidth]{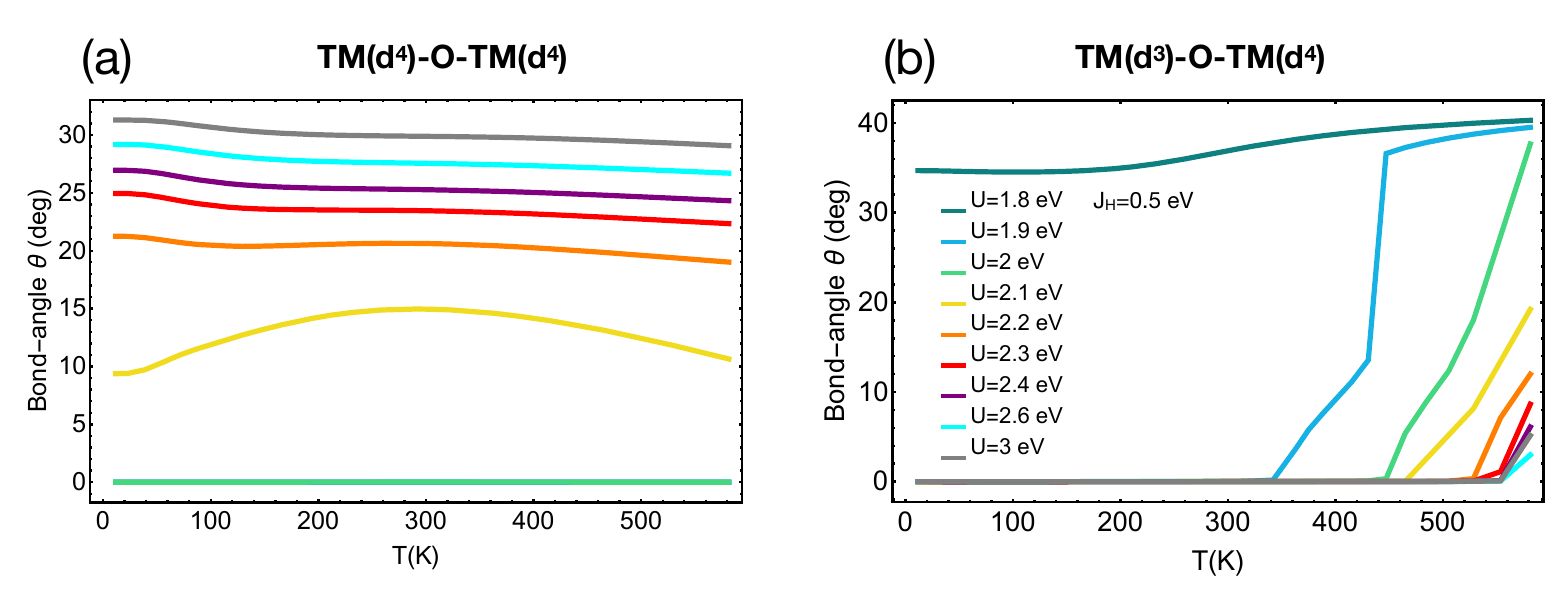}
\caption{
Temperature dependence of the bond angles for
(a) TM($d^4$)-O-TM($d^4$) and
(b) TM($d^3$)-O-TM($d^4$) bonds.
The curves have been obtained fixing $J_H=0.5\,$eV, for different values of $U$, as reported in the legend. The other parameters, measured in eV, are
$V_{pd\pi}=1.3$, $V_{pd\sigma}=1.6$, $\delta=0.25$, $\delta_{ort}=0.09$, $\lambda$=0.075, the latter being the spin-orbit coupling.}
\label{d3-d4}
\end{figure}

A similar pattern emerges for the TM($d^3$)-O-TM($d^4$) bond configuration. Notably, varying electronic parameters within the defined physical regions does not markedly alter the phase diagram's structure. This trend can be understood qualitatively by considering the dependence of $p$-$d$ hybridization on the TM-O-TM bond angle. For instance, a large $V_{pd\pi}$ favors $\theta\simeq 0$ to optimize $p$-$d$ charge transfer and kinetic energy. Conversely, significant $V_{pd\sigma}$ prompts a deviation from $\theta\simeq 0$, enhancing kinetic energy through $p$-$d$ hopping processes. Next, we explore the thermal evolution of the optimal bond angle. We identify, at each temperature, the free energy minimum concerning the TM-O-TM bond angle. Initially, we select a $d$-$p$ hybridization regime where the ground state implies a nonzero bond angle, indicating low-temperature bond distortion. We then track the temperature-dependent evolution of $\theta_{\rm opt}$, optimizing free energy by adjusting the Coulomb interaction $U$ and the Hund's coupling $J_H$.
From Fig.~\ref{d3-d4}(a), we see that in the case of TM($d^4$)-O-TM($d^4$) the decrease of $U$ induces a transition from a regime where the bond angle remains unaffected by temperature changes to one where the bond angle is appreciably affected by temperature variations. Moreover, it is also found that Hund's coupling affects the critical $U$ value for transitioning to an undistorted bond angle, favoring higher $J_H$ values~\cite{Brzezicki23}. Additionally, it influences the temperature at which a bond angle variation is observed, underscoring the link between spin, orbital correlations, and thermal bond angle change. In Fig.~\ref{d3-d4}(b), we explore the hybrid TM($d^3$)-O-TM($d^4$) bond, where substituting a $d^4$ TM atom with a $d^3$ impurity alters electron occupation at one TM site. Here, the optimal bond angle exhibits a distinct trend from the TM($d^4$)-O-TM($d^4$) bond. With increasing $U$, there is a noticeable tendency for the bond angle to decrease towards a less distorted configuration. Conversely, as $U$ increases and $J_H$ decreases, the bond angle increases with decreasing temperature. As can be seen from the schematic representation in Fig.~\ref{bond}(c), an increase in the bond angle with decreasing temperature corresponds to a reduction in the square area of the TM-O lattice, which in turn leads to a reduction in the unit cell volume, thereby explaining the observed NTE.

In summary, this study unveils essential mechanisms for achieving NTE effects. Anisotropic orbital correlations, driven by electron interactions, play a significant role in promoting a ground state conducive to NTE. At low temperatures, strong spin-orbital correlations lead to less distorted bond directions, while thermal fluctuations weaken these correlations, favoring more distorted bonds. Thermal disorder further impacts the spin channel, generating non-collinear magnetic correlations that energetically support distorted bonds. These conditions culminate in a reduction of the optimal bond angle with increasing temperature, manifesting NTE effects.

Integrating an electron-lattice interaction could potentially moderate the pronounced bond distortions observed in a purely electronic framework, facilitating smoother transitions between varying bond angles influenced by $p$-$d$ hybridization dynamics. One conceivable strategy might entail employing a phenomenological model featuring a quadratic electron-lattice potential, where the coupling strength correlates with the optimal angle obtained from minimizing the electronic system's free energy. This assumption is grounded in the notion that the lattice acts akin to a restoring force, and its impact, mediated via electron-phonon coupling, may be contingent upon the optimal angle determined by electronic considerations. Further insights into this methodology are elucidated in Ref.~\cite{Brzezicki23}.

\section{Electric field effects}

Designing emergent phases and their physical properties via electric gating is at the forefront of modern condensed matter physics. Recently, many observations in a variety of correlated materials have indicated that the application of an electric field, apart from leading to a dielectric breakdown with metal-insulator transition (MIT) in quasi-static conditions, can yield novel phase transitions in the insulating regime which cannot otherwise be accessible~\cite{Cao2018, Stojchevska2014, Afanasiev2019}. Ca$_2$RuO$_4$ is a well-established example of such systems. It exhibits a complex phase diagram and a strong susceptibility to phase changes in the insulating state, which is extremely sensitive to different kinds of external perturbations. As a consequence, the MIT which characterizes its behavior can be tuned, besides by heating, through various stimuli, such as chemical substitution~\cite{Porter2022}, pressure~\cite{Steffens2005,Nakamura2002}, epitaxial strain~\cite{Ricco2018}. Recently, it has been demonstrated that at room temperature the equilibrium insulating phase of Ca$_2$RuO$_4$ can be transformed in a metallic one by the application of an electric field, through current and optical pulses.~\cite{Nakamura2013,Cirillo19, Zhang2019}. {As shown in Ref.\cite{Nakamura2013}, the phase transition is induced by an applied voltage $V\simeq 0.8\,$V which corresponds to an electric field $E\simeq$ 40\,V/cm. This value is unusually small compared to the Mott energy gap, being about two orders of magnitude below a typical break-through field of a Mott insulator~\cite{Taguchi2000}. }However, the electronic and structural signatures of this electric current-induced phase are substantially different from those of the conventional metallic phases obtained by temperature, pressure, or doping~\cite{Cirillo19, Zhang2019, Bertinshaw19}. Experimentally it was found that the MIT under current is not due to a breakdown of the Mott state, but rather stems from the formation of in-gap states, with the Mott state that is only reorganized~\cite{Curcio23}. Moreover, recent nano-resolved infrared probe studies have detected a non-trivial morphology in current-driven metallic domains, growing from one electrode and forming specific patterns. A salient feature of the phase boundaries is its relationship with specific crystallographic orientations and its arrangement in micrometric stripes, which hints to a strong correlation between crystallographic and electronic degrees of freedom at the electric current-induced MIT~\cite{Zhang2019}. Notably, the capability to induce a nanoscale stripe domain through electric-field quenching in Ca$_2$RuO$_4$~\cite{Gauquelin23} was also demonstrated. While the zero-current MIT is understood as a consequence of  multi-band physics, Coulomb interaction, Hund's coupling and the non-trivial interplay with structural distortions~\cite{Sutter17}, the mechanism by which the current induces the MIT is not yet established.

In this Section we analyze the impact on the electronic degrees of freedom of an applied electric field and which states arise in non-equilibrium steady state conditions.
As mentioned in Section 2, in Ca$_2$RuO$_4$ the low-temperature insulating phase occurs accompanied by a strong first-order structural transition, characterized by the shrinking of the $c$ axis and a concomitant elongation of the in-plane parameters. Such structural deformations correspond to a transition from elongated RuO$_6$ octahedra at high temperature to flattened octahedra in the insulating state, but also lead to the increase of the tilting and rotation of the RuO$_6$ octahedra along the orthorhombic $b$ direction~\cite{Braden98,Friedt01,Steffens2005}. This latter distortion seems to pin the magnetic moment parallel to the $b$ axis~\cite{Porter2022}.
The structural transition involves an orbital reconstruction leading to a preferential occupation of the 4$d$ states with in-plane character ($xy$) with respect to the out-of-plane states ($xz, yz$). At lower temperatures, this redistribution turns into an orbitally ordered state, and the antiferromagnetic and orbital correlated phases at the equilibrium can be qualitatively understood in the framework of spin-orbital exchange mechanisms for which ferro-orbital type of configurations come along with the antiferromagnetism.

We demonstrate that the spin-orbital correlations are dramatically affected by the application of the electric field. The resulting scenario is that the equilibrium configurations can locally evolve into states with strong orbital fluctuations and destruction of the orbital unbalance. We foresee that in this regime, the magnetic correlations may lead to competition/frustration in the time evolution between antiferromagnetism and ferromagnetism, with a tendency to suppress the dominant antiferromagnetic correlations. The resulting time evolution thus clearly indicates that in the Mott state the spin-orbital correlations can be modulated in amplitude without requiring structural changes. This result goes beyond the standard equilibrium view of spatially ordered patterns for both spin orientations and orbital occupations, thus opening novel pathways for switching spin and orbital orders in the Mott phase through nonequilibrium electrical stimuli.

\subsection{Method}
We employ the model Hamiltonian describing the Ru and O bands close to the Fermi level for the itinerant electrons within the ruthenium-oxygen plane, described by the Hamiltonian in Eq.(\ref{Ham_NTE}) and detailed in the Appendix.
We apply this model on a finite cluster comprising two ruthenium ions, Ru$_1$ and Ru$_2$, linked by a single oxygen atom O, under the influence of an external electric field, so that 
\begin{eqnarray}
H_{\rm TM} & = & H_{\rm Ru_1} + H_{\rm Ru_2} \\
H_{\rm TM-O} & = & H_{\rm Ru_1-O}+H_{\rm Ru_2-O} \, . \label{H_RuO}
\end{eqnarray}
Formally, we introduce the effect of an external electric field through the Peierls substitution, which modifies the the hopping Hamiltonian in Eq.(\ref{H_RuO}) via the Peierls phase factor,
\begin{eqnarray}
t_{d_{\alpha},p_{\beta}}(t)=t_{d_{\alpha},p_{\beta}} \exp\left[-i\,\frac{e}{\hbar} \int_{r_{\rm Ru}}^{r_{\rm O}} {\bf{A(t)}} d{\bf r} \right] \, ,
\label{Peierls}
\end{eqnarray} 
where we indicate with $r_{\rm O}$ and $r_{\rm Ru}$ the position of the O and Ru ions, and with $e$ and $\hbar$ the electron charge and the Planck constant, respectively. It should be noted that the vector potential in Equation (\ref{Peierls}) is connected to the electric field through the relation $\bf{E}(t)=-\partial_{t}{\bf A(t)}$. 
Specifically, we consider a constant electric field oriented along the Ru-O-Ru axis, thus we utilize a time-dependent scalar potential described by $A(t)= E t$. 

Introducing the term in Eq.(\ref{Peierls}) necessitates solving the time-dependent Schrödinger equation, $i \hbar \frac{\partial}{\partial t} |\Psi(t)\rangle = H(t) |\Psi(t)\rangle$, which governs the temporal evolution of the quantum system at absolute zero. To achieve this, we implement the following approach. Initially, we identify the ground state of the Hamiltonian through exact diagonalization. Subsequently, we evaluate the temporal evolution of the many-body ground state using the Cranck-Nicholson method, which ensures a unitary time evolution, where the wave function at a later time is represented within an infinitesimal time interval as
\begin{eqnarray}
\Psi(t+\Delta t)\rangle &&=\exp[-i \hbar^{-1} \int_{t}^{t+\Delta t} H(t) dt] |\Psi(t)\rangle \nonumber \\ 
&& \approx \frac{[1-i \frac{\Delta t}{\hbar} H(t+ \Delta t)]}{1+i \frac{\Delta t}{\hbar} H(t+ \Delta t)] }  |\Psi(t)\rangle \,.
\end{eqnarray}
We adopt a time increment of $dt=1.0\times 10^{-2} \hbar/t^{0}{p,d}$, where $t^{0}{p,d}$ denotes the magnitude of the $p-d$ $\pi$-hybridization hopping parameter. This time step is sufficiently small to ensure the solution's convergence. Subsequently, we track the time evolution of the ground state, focusing particularly on the on-site orbital populations of the $d$-orbitals. We achieve this by computing the time-dependent expectation value of the electron density $n_{xy}$ in the d$_{xy}$ orbital and the averaged electron density in the d$_{xz}$ and d$_{yz}$ orbitals, expressed as $\frac{1}{2}(n_{xz}+n_{yz})$.

\subsection{Discussion of the results}

To evaluate the electric field driven properties of the examined multi-orbital model, we determine the time dependent evolution of the ground state by discretizing the time interval. 
Since we are dealing with Ca$_2$RuO$_4$ system, we fix the electronic parameters in the regime that is better matching with the monolayer ruthenate compound. There, the octahedra become flat below the structural transition temperature, hence the crystal field splitting $\Delta_{\rm CF}$ is negative and, according to first principle calculations or estimates employed to reproduce the resonant inelastic x-ray and neutron scattering spectra, its amplitude is in the range $\sim$ [200-300] meV.
Furthermore, it is useful to point out that material-specific parameters such as $\lambda=0.075$ eV ($\lambda$ is the spin-orbit coupling, see the Appendix), $U$ in the range [2.0,2.2] eV, and $J_{H}$ [0.35, 0.5] are used as benchmarks for the study. Comparable values for $\Delta_{\rm CF}$, $U$ and $J_{\rm H}$ have been applied in computations of the electronic spectra in Ca$_2$RuO$_4$, and the ratio $g=\Delta_{\rm CF}/(2 \lambda)$ is generally assumed to fall within a range approximately equal to [1.5, 2] for simulating the spin excitations detected via neutron scattering. For the hopping integrals, it is assumed that the fundamental $p$-$d$ hopping amplitudes in the tetragonal ($\theta=0$) symmetry have the value $t^{0}_{p,d}$=1.5 eV.

\begin{figure*}[htbp]
\centering
\includegraphics[width=14cm]{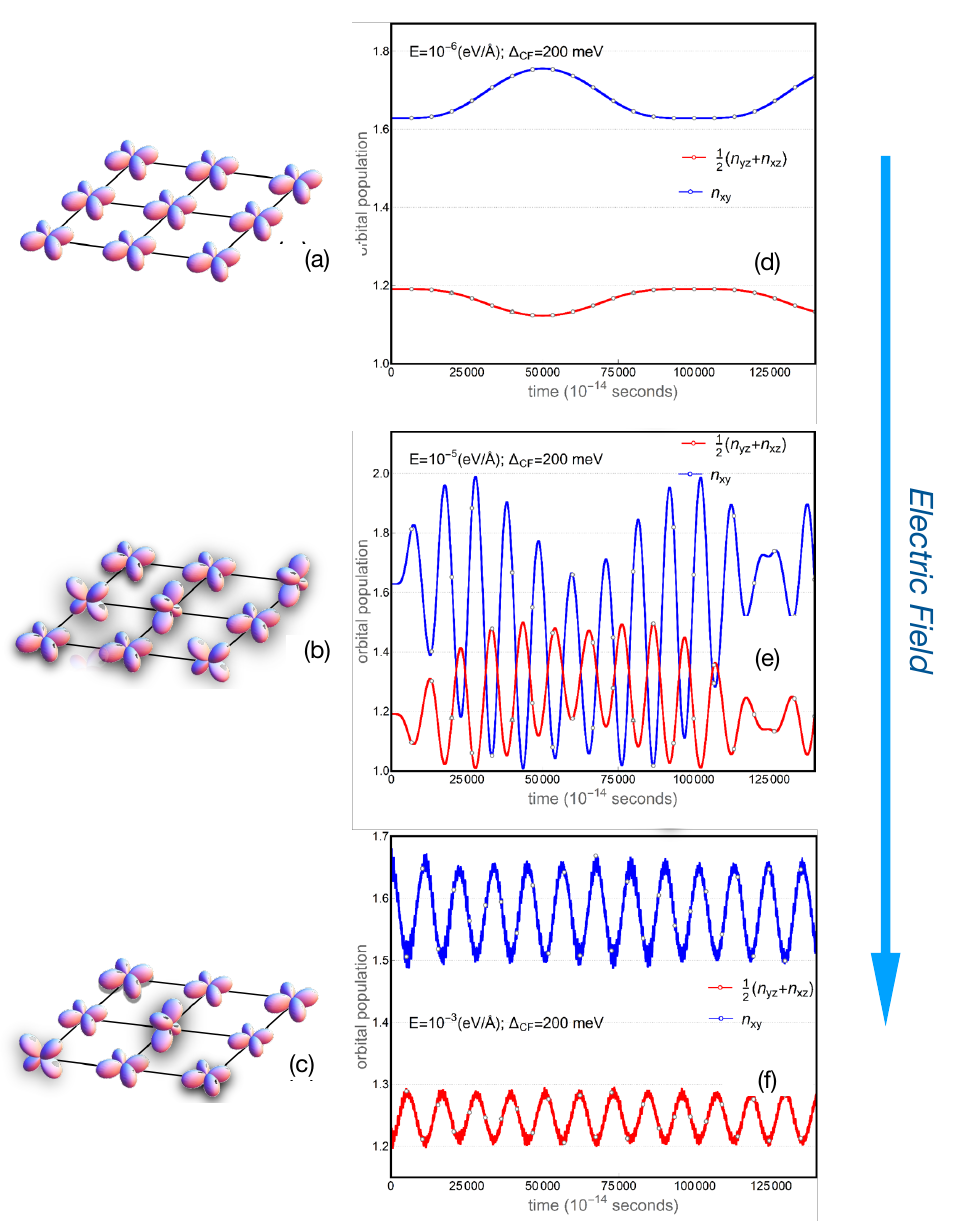}
\caption{Evolution of the orbital occupation at the Ru site for a given tetragonal distortion corresponding to $\Delta_{\rm CF}=200$ meV and electric field ampltitudes (d) $E=10^{-6}$ eV/\AA, (e) $E=10^{-5}$ eV/\AA, (f) $E=10^{-3}$ eV/\AA \; corresponding to dominant configurations with strong orbital unbalance (a), orbital collapse (b) and intermediate orbital modulation (c), respectively. We demonstrate that a weak electric field can completely destroy the orbital order characterized by dominant $xy$ electron density. Electronic parameters are $J_{\rm H}=0.4$ eV, $U=2.0$ eV, and the Ru-O-Ru bond angle is $\theta=10^{o}$.}
\label{FigElField}
\end{figure*}

At zero electric field the ground state is antiferromagnetic with anisotropic magnetic correlations and ferro-type orbital pattern with the doublon mainly occupying the d$_{xy}$ state. Switching of the electric field can drive a tunneling between different spin-orbital quantum states in time thus altering the magnetic and orbtial character of the driven state. We report in Fig.~\ref{FigElField} the time dependent view of the orbital occupation at the Ru site upon the application of electric fields with growing amplitude. Starting from a configuration having an orbital pattern with $(n_{xy},n_z)=(1.6,1.2)$ a tiny perturbation of the electric (i.e. $E=10^{-6}\,$eV/\AA), sets orbital modulation on a time scale of the order of 0.1 ns, with an increasing of the orbital unbalance. This means that for such values of the electric field drive the most favorable quantum tunneling process are with states that prevent that doublon to occupy the $(xz,yz)$ configurations. The increase of the electric field in magnitude activates a completely different time dependence for the ground state. Indeed, the orbital fluctuations grow in intensity and the quantum system evolves from configuration with almost fully double occupied d$_{xy}$ to others with a complete suppression of the orbital ordering with a striking orbital population inversion of the $xy$ and $z$ configurations. Although we cannot reach a steady state due to the finite size of the investigated cluster, the regime of orbital collapse persists in a window of time of the order of nanosecond thus indicating to be a robust feature of the out-of-equilibrium phase. In the regime of large electric field, the orbital fluctuations are reduced in intensity. The time evolution still exhibits a tendency to suppress the orbital unbalance between the $xy$ and $z$ configurations. However, in this regime one does not find evidences of a full suppression of the orbital ordering. The behavior is markedly distinct from that one of the weak electric field because the orbital unbalance is reduced and the orbital fluctuations are faster in time occurring on a scale of picoseconds.  
The resulting scenario is that of an orbital hardening-collapse-softening changeover of the local orbital ordering as a function of the intensity of the applied electric field. 
Such behavior is also observed for larger values of the crystal field potential, $\Delta_{CF}$, until a critical threshold above which the regime of orbital collapse disappears and the non-equilibrium quantum evolution moves from the hardening to the softening of the local orbital fluctuations.

In conclusion, we find that the ground state can be fully electrically tuned in a model system that is suited for the experimental case of Ca$_2$RuO$_4$ Mott insulator. A weak electric field is able to independently destroy  the orbital order without substantial changes in the crystal structure. Such observation sets out a clear difference between thermal and electric driven phases in Ca$_2$RuO$_4$. The theoretical simulations supports a scenario where spin-orbital correlations at the equilibrium can be completely reconstructed upon the application of weak electric field. The ground state configuration evolves in time through states with strong orbital fluctuations and destruction of the orbital polarization, but without requiring structural changes to less compressed RuO$_6$ octahedra, which at equilibrium are normally associated to the orbitally unpolarized state. In this regime, the magnetic correlations are also expected to be substantially modified, pointing to a sort of magnetic frustration in the time evolution which may also have a role in the observed suppression of antiferromagnetism. Finally, we point out that recently, the effect of a quenched electric field was studied, both experimentally and theoretically~\cite{Gauquelin23,Cuono22}. 
Supported by ab-initio calculations, it has been demonstrated that the application of an electric field may support alternated stripe patterns with longer/shorter $c$ axis, where the Ru atoms are displaced from the centrosymmetric positions~\cite{Gauquelin23}. 

\section{Conclusions and perspectives}

The subtle competition between multi-orbital effects and lattice degrees of freedom in Ca$_2$RuO$_4$ offers the possibility to trigger exotic quantum states. Here, we have analysed recent achievements about these new quantum phases focusing on altermagnetism, negative thermal expansion and electrically tuned non-conventional emerging ground states.

Altermagnetism is a recently discovered magnetically ordered phase, different from antiferromagnetism and ferromagnetism, where the electronic charge of spin-up (down) atoms ought to be transformed into spin-down (up) atoms not through translations, inversions, or their combinations, but rather via the utilization of roto-translations or mirrors~\cite{Smejkal22,Smejkal22b}. This magnetic phase is at the center of today's debate and there are numerous works aimed at discovering new altermagnets. These systems present spin-splitting in the band structure like ferromagnets, but they have no net magnetization in the real space, like antiferromagnets. Therefore they have similarities with both the phases, and since antiferromagnetism has been found in a wider range of materials than ferromagnetism and appears at higher temperatures, altermagnets offer a new avenue for numerous applications. The altermagnets are useful in spintronics~\cite{Shao21}, spincaloritronics~\cite{Zhou24}, Josephson junctions~\cite{Ouassou23} and they can bring to highly efficient spin-current generation~\cite{Hernandez21}. 

It has been shown, by means of density functional theory calculations, that Ca$_2$RuO$_4$ is a orbital selective altermagnet~\cite{Cuono23}, where the non relativistic spin-splitting only appear in the d$_{xz}$-d$_{yz}$ bands while the splitting is suppressed in the d$_{xy}$ band subsector. This is due to the different character of these orbitals, while  d$_{xy}$ are quasi-two-dimensional because they have negligible hybridizations between the layers, the d$_{xz}$-d$_{yz}$ bands are 3D in the sense that they present hybridization along $z$. The altermagnetic spin-splitting was shown to be present along some lines of the Brillouin zone. While the splitting is maximum along $\Gamma$-R and $\Gamma$-U, it survives also in all the Brillouin zone except in the high-symmetry directions where one of the k-component is zero or on the zone boundaries. The discovery of altermagnetism in Ca$_2$RuO$_4$ is a new step in the analysis of compounds that can exhibit these fascinating properties really important both from a theoretical point of view and from an application point of view. Due to orbital-selective antiferromagnetism,  the anomalous Hall effect can manifest exclusively in n-doped samples of Ca$_2$RuO$_4$, with no significant altermagnetism observed in the highest valence bands.

Referring to NTE, our study identifies key mechanisms crucial for achieving NTE effects. Anisotropy in orbital correlations, driven by electron correlations, is highlighted as significant. This anisotropy leads to preferential orbital occupation, fostering a ground state prone to NTE effects. At low temperatures, strong spin-orbital correlations drive bond directions to be less distorted. Conversely, thermal fluctuations weaken these correlations, favoring more distorted bonds. Thermal disorder also affects the spin channel, generating non-collinear magnetic correlations that energetically support distorted bonds. These conditions lead to a decrease in the optimal bond angle with increasing temperature, resulting in NTE effects.

While NTE materials are still relatively uncommon, their unique properties open up possibilities for practical applications. For instance, NTE materials could be incorporated into composites or structures to counteract the thermal expansion of other components, thereby enhancing dimensional stability. This could find applications in industries where precise dimensions are critical, such as aerospace and optics. Despite the promising aspects of NTE, challenges remain in synthesizing materials with controlled NTE properties and understanding the underlying physics at a fundamental level. Future research directions may involve exploring novel NTE materials, refining our understanding of the mechanisms, and developing practical applications that leverage this intriguing property. Moreover, a relevant possible extension of the approach presented here is the inclusion of the electron-phonon interaction in the model, to be considered on an equal footing with the electronic correlations. Incorporating an electron-lattice potential could help regulate the extreme bond distortions seen in a purely electronic model and facilitate smoother transitions between large and small bond angles across the phase space influenced by $p$-$d$ hybridization hoppings. A possible approach may involve a phenomenological model with a quadratic electron-lattice potential, where the coupling constant is proportional to the optimal angle derived from minimizing the free energy of the electronic system. This assumption is supported by the idea that the lattice behaves as a restoring force, and the strength of its influence, mediated by electron-phonon coupling, might depend on the optimal angle dictated by electronic factors. Details on this approach can be found in Ref.~\cite{Brzezicki23}.

Furthermore, the interplay between electronic correlations, spin-orbit coupling, and structural distortions triggers non-standard Mott physics and unconventional spin-orbital ordering, potentially resulting in a high sensitivity to external stimuli like electric current or electric field.
Here, we have considered the effect of an applied electric field on Ca$_2$RuO$_4$ and found that spin-orbital correlations can be fully electrically driven. A weak electric field is able to destroy the orbital order without substantial changes in the crystal structure. This result goes beyond the standard equilibrium view of spatially ordered patterns for both spin orientations and orbital occupations, thus opening novel paths for switching spin and orbital orders in a Mott phase through non-equilibrium electrical stimuli.

\section{Appendix}

\noindent Here we specify the form of the three terms $H_{\mathrm{TM}}$, $H_{\rm O}$ and $H_{{\rm TM}-{\rm O}}$ entering the Hamiltonian (\ref{Ham_NTE}).

The local Hamiltonian $H_{\mathrm{TM}}$ at the $i$-th TM site is given by 
\cite{Cuoco06a,Cuoco06b,Forte10} 
\begin{equation}
H_{\mathrm{TM}}=H_{e-e}+H_{\mathrm{CF}}+H_{\mathrm{SOC}} \, 
\end{equation}
where 
\begin{eqnarray}
H_{e-e}&=&
U\sum\limits_{\alpha} n_{i\alpha\uparrow}n_{i\alpha\downarrow}
+J_{H}\sum\limits_{\alpha\ne\beta}
d_{i\alpha\uparrow}^{\dagger}d_{i\alpha\downarrow}^{\dagger}
d_{i\beta\uparrow}^{}d_{i\beta\downarrow}^{}  \nonumber\\
&+&\left(U^{'}-\frac{ J_{\mathrm{H}}}{2}\right)
\sum\limits_{\alpha <\beta }n_{i\alpha }n_{i\beta }
-2J_{\mathrm{H}}\sum\limits_{\alpha<\beta}
({\bf{S}})_{i\alpha}\!\cdot\!({\bf{S}})_{i\beta} \label{He-e}\\
\nonumber \\
H_{\mathrm{CF}} &=&\varepsilon_{xy} n_{i,d_{xy}}+ \varepsilon_{\gamma z}\left(n_{i,d_{xz}}+ n_{i,d_{yz}}\right)\, \\  \nonumber\\
H_{\mathrm{SOC}}&=&\lambda \sum\limits_{\alpha ,\sigma }
\sum_{\beta ,\sigma^{^{\prime }}}\,                        
d_{i\alpha \sigma }^{\dagger }\,({\bf{L}})_{\alpha\beta}
\cdot ({\bf{S}})_{\sigma \sigma ^{^{\prime }}}\,
d_{i\beta\sigma ^{^{\prime }}} \, .
\end{eqnarray}
Here $H_{e-e}$ includes the complete Coulomb interaction projected on the $t_{2g}$ electrons~\cite{PhSS_14}, $H_{\rm CF}$ is the tetragonal crystal field potential and $H_{\rm SOC}$ is the spin-orbit coupling. The indices $\{\alpha,\beta\}$ denote the three orbitals in the $t_{2g}$ sector, i.e., \mbox{$\alpha,\beta\in\{d_{xy},d_{xz},d_{yz}\}$,} $d_{i\alpha\sigma}^{\dagger}$ represents the creation operator of an electron with spin $\sigma$ at the site $i$ in the orbital $\alpha$. The operator $\bf{L}$ is the angular momentum operator in the $t_{2g}$ subspace and $\bf{S}=\frac{1}{2}\pmb{\sigma}$ is the spin operator expressed through the Pauli matrices.

The local Hamiltonian $H_{\rm O}$ at the $j$-th O site only includes the on-site energy terms for $p_x, p_y, p_z$ orbitals and it is given by:
\begin{equation}
H_{\mathrm{O}} =\varepsilon _{x}n_{j,p_x}+\varepsilon _{y}
n_{j,p_y}+\varepsilon_{z}n_{j,p_z} \, .
\end{equation}

Finally, for the TM-O hopping term $H_{{\rm TM}-{\rm O}}$ we assume that for a generic bond connecting the TM to the O atoms along the $x$-direction, the $p$-$d$ hybridization includes all the allowed hopping terms between electrons belonging to $p$ and $d$ orbitals, having the general form
\begin{equation}
H_{{\rm TM}-{\rm O}}= t_{d_{\alpha},p_{\tau}}\left(
d_{i,\alpha}^{\dagger }\,p_{i+a_{x},\tau}+H.c.\right)\, .
\end{equation}
We consider the TM-O allowed hoppings according to the Slater-Koster rules~\cite{Slater1954,Brzezicki2015} for a given bond connecting a TM to an oxygen atom along a given symmetry direction, e.g. the $x$ axis. They are given by
\begin{eqnarray}
t_{d_{xy},p_x}&&=
\sqrt{3} \, n_x^2 \,  n_y V_{pd\sigma}+n_y (1-2 n_x^2) \,  V_{pd\pi}, \nonumber \\
t_{d_{xy},p_y}&&=
\sqrt{3} \,  n_y^2 \, n_x V_{pd\sigma}+n_x (1-2 n_y^2)\,  V_{pd\pi}, \\
t_{d_{xz},p_z}&&=n_x \, V_{pd\sigma}, \nonumber \\
t_{d_{yz},p_z}&&=n_y \, V_{pd\pi}, \nonumber
\end{eqnarray}
with $n_x=\cos \theta$ and $n_y=\sin \theta$.
In a similar way, one can write down the other hybridization
terms along the $y$-direction for the other TM-O bonds.

\end{document}